\documentclass[aps,prd,amsmath, showpacs, floatfix]{revtex4}
\usepackage{graphicx}
\usepackage{bm}
\let\l=\left
\let\r=\right
\def\be{\begin{equation}}
\def\ee{\end{equation}}
\def\bea{\begin{eqnarray}}
\def\eea{\end{eqnarray}}

\begin{document}

\title{Numerical simulation of oscillatons: extracting the radiating tail}

\author{Philippe Grandcl\'ement}
\affiliation{LUTH, CNRS-UMR 8102, Observatoire de Paris-Meudon, place Jules Janssen, 92195 Meudon Cedex, FRANCE}
\author{Gyula Fodor}
\affiliation{MTA RMKI, H-1525 Budapest 114, P.O.Box 49, Hungary}
\author{P\'eter Forg\'acs}
\affiliation{MTA RMKI, H-1525 Budapest 114, P.O.Box 49, Hungary}
\affiliation{LMPT, CNRS-UMR 6083, Universit\'e de Tours, Parc de Grandmont, 37200 Tours, FRANCE}
%
%
%
\date{\today}

\begin{abstract}
Spherically symmetric, {\sl time-periodic oscillatons} -- solutions of
the Einstein-Klein-Gordon system (a massive scalar field coupled to gravity) with a spatially localized core -- are investigated by very precise numerical techniques based on spectral methods. In particular the amplitude of their standing-wave tail is determined.
It is found that the amplitude of the oscillating tail is very small, but non-vanishing
for the range of frequencies considered.
It follows that exactly time-periodic oscillatons are not truly localized, and they can be pictured loosely as consisting of a well (exponentially) localized nonsingular core and an oscillating tail making the total mass infinite. Finite mass physical oscillatons with a well localized core -- solutions of the Cauchy-problem with suitable initial conditions -- are only approximately time-periodic. They are continuously losing their mass because the scalar field radiates to infinity. Their core and radiative tail is well approximated by that of
time-periodic oscillatons.
Moreover the mass loss rate of physical oscillatons is estimated from the numerical data
and a semi-empirical formula is deduced.
The numerical results are in agreement with those obtained analytically in the limit of small amplitude time-periodic oscillatons.

\end{abstract}

\maketitle

\section{Introduction}

Oscillating soliton stars with spherical symmetry were first observed in numerical simulations performed by Seidel and Suen in \cite{SeideS91}. They considered a free, real {\sl massive} scalar field
(of mass $m$) coupled to gravity. Starting from general initial data, it was found that the system could settle in a spatially localized and apparently time-periodic and stable state.
Such spatially localized, oscillating solutions are referred to as oscillatons.
There seems to be at least a one parameter family of oscillatons, labeled by their frequency, $\omega<m$.
The numerical simulations were consistent with oscillatons being time-periodic, with a constant frequency. Assuming exact periodicity the structure of such solutions was then investigated by means of a Fourier decomposition of the various fields describing the system.
At that time, it was not clear if oscillatons were truly periodic and Seidel and Suen left open the possibility of the appearance of a secular change of their frequency. The same authors studied the formation of oscillatons in a scenario involving gravitational cooling in the companion paper \cite{SeideS94}, making them of possibly great physical importance.  Oscillatons appear to be good candidates for dark matter in our Universe \cite{Alcubgalactic,Susperregi,Guzman1,Hernandez,Guzman3,Bernal}.

Some years later, this subject was re-investigated in a series of papers \cite{Lopez02, LopezMB02, AlcubBGMNU03}. Several aspects of oscillatons were studied, like their structure, dynamics, stability and more. In particular it was found that oscillatons reached a maximum mass around an oscillation frequency of $\omega_{\text{min}}\approx 0.86m$. It is expected that solutions would become unstable below this frequency,
therefore the frequency range of oscillatons is expected to be $\omega_{\text{min}}<\omega<m$.
The possible existence of a secular change of the frequency of oscillatons induced by (scalar) radiation has not been discussed, however, in those papers, as a matter of fact it has been simply assumed that the solutions are truly periodic.

The physically motivated, important question concerning the possible radiation loss of oscillatons, has been first considered by Don N.\ Page \cite{Page}, who has pointed out
that due to scalar radiation oscillatons are bound to lose mass.
Page has estimated the amplitude of the outgoing scalar wave
and derived a formula for their mass-loss, which turns out to be
rather small even on cosmological time-scales.
The subject of longevity of oscillatons has been taken up again in a recent paper \cite{FodorFM10}. This study has also been performed in the limit where the amplitude of the scalar field is small, when a perturbative analytic approach is feasible,
and the structure of oscillatons was investigated in detail.
In the limit of small amplitudes, oscillatons correspond to solutions with frequencies $\omega \rightarrow m$.
The fact that oscillatons are continuously losing mass due to scalar radiation,
leading to secular changes in their frequency has been confirmed.
A formula for the mass loss of oscillatons in spatial dimensions $2<D<6$ has been given.
Although the quantitative result of Ref.\ \cite{FodorFM10} for the mass loss rate differs significantly from that of Ref.\ \cite{Page}, there is qualitative agreement on its extreme smallness as compared to the total energy of the configurations. This explains why \emph{physical oscillatons}, evolving from finite energy initial data, are practically indistinguishable from \emph{exactly time-periodic oscillatons}, which have an extremely small amplitude standing wave tail, and also why this effect has been largely ignored by previous numerical work.

In various field theories containing massive scalars, very similar objects, oscillons or pulsons have been numerically observed and analytically investigated mostly in the limit of small amplitudes \cite{BogMak2,CopelGM95,Honda,Hindmarsh-Salmi06,SafTra,Farhi05,koutv,Graham07b,AminShirkoff,Herzberg,fggirs,GleiserGraham}. Oscillatons under the name of ``gravipulsons'' have been studied in the recent paper \cite{Koutvitsky-Maslov} in EKG theory with a logarithmic scalar self-interaction analytically without assuming a small amplitude limit 

The goal of this paper is to investigate exactly time-periodic oscillatons in $3$ dimensional EKG theory and generalize some of the results of Ref. \cite{FodorFM10} without
using the small amplitude limit.
The asymptotic oscillating tails of the fields are determined by special high precision numerical techniques. The extreme smallness of the oscillatory tail of an oscillaton with respect to its central amplitude \cite{FodorFM10} explains why it has been missed in previous numerical work. We show that nevertheless this tail can be computed by making use of very precise numerical techniques. In this context, the field equations are solved using spectral methods, where the variables are approximated as finite sums of known functions called the basis functions. This is done for both space and time. The solution of the numerical system is performed by using the spectral solver Kadath \cite{kadath, Grand10} which enables the use of spectral methods in a wide range of problems arising in theoretical physics. The oscillating tails of the scalar fields are obtained by using a similar approach as the one used in Ref.\ \cite{FodorFGR06} in the case of oscillons in a self-interacting single scalar field theory in $3$ dimensions.
A detailed comparison of the numerical results with those obtained in the small amplitude limit
shows remarkable agreement and coherence. The core of the oscillaton is approximated by the expansion in the small amplitude limit to good precision even for not too small values of the amplitude.
This expansion significantly underestimates, however, the magnitude of the oscillating tail
which depends in an essentially non-analytic way from the amplitude.

The paper is organized as follows. In Sec. \ref{s:model} the equations describing a scalar field coupled to gravity are presented, along with the decomposition in modes used to obtain periodic and weakly localized solutions. The expected asymptotic behaviors of the various fields are also studied. Section \ref{s:numerical} is devoted to the presentation of the numerical techniques. The use of the library Kadath and the way solutions are matched at the outer end of the computational domain are explained in some detail. Numerical results are shown in Sec. \ref{s:results} along with many numerical tests that validate the overall procedure. The existence of an oscillatory tail and comparison with previous work are discussed. The comparison with results from the small-amplitude expansion is shown in Sec. \ref{s:small}. The mass loss rate of oscillatons is determined in Sec. \ref{s:longevity}.

\section{The model}
\label{s:model}
\subsection{The equations}
\label{ss:equations}

One considers a real scalar field $\Phi$, coupled to gravity. The stress-energy tensor is given by
\be
T_{\mu\nu} = \Phi_{,\mu} \Phi_{,\nu} - g_{\mu\nu} \l[\frac{1}{2} \Phi_{,\alpha} \Phi^{,\alpha} + U(\Phi)\r],
\ee
where $U(\Phi)$ is the potential of self-interaction.

We are interested in finding spherically symmetric configurations in 3+1 dimensions. We can chose the metric to take the following form :
\be
{\rm d}s^2 = - A {\rm d}t^2 + B \l({\rm d}r^2 + r^2 {\rm d}\Omega\r),
\ee
where $\Omega$ is the solid angle. The unknown functions $A$ and $B$ depend solely on $r$ and $t$, and this system of coordinates may be described as quasi-isotropic.

In order to get rid of the $8\pi$ factors (see Sec. II-A of \cite{FodorFM10}) we rescale the scalar field and the potential as
\begin{equation}
\Phi\to\frac{1}{\sqrt{8\pi}}\,\Phi \ , \quad U(\Phi)\to \frac{1}{8\pi}\,U(\Phi) \ . \label{e:scscale}
\end{equation}
Under these assumptions, Einstein's equations $G_{\mu\nu} = 8\pi T_{\mu\nu}$ are written as :
\bea
\label{e:gtt}
2 \l[ \frac{3}{4} \l(\frac {B_{,t}}{B}\r)^2 - \frac{A B_{,rr}}{B^2} - 2 \frac{A}{B^2}\frac{B_{,r}}{r}
+ \l(\frac{3}{4}\r) \frac{A\l(B_{,r}\r)^2}{B^3}\r] &=&
\l(\Phi_{,t}\r)^2 + \frac{A}{B}  \l(\Phi_{,r}\r)^2 + 2AU(\Phi) \\
\label{e:grr}
2 \l[ \frac{B_{,r}}{Br} + \frac{1}{4}\l(\frac{B_{,r}}{B}\r)^2 + \frac{A_{,r}}{Ar} + \frac{1}{2} \frac{B_{,r}}{B} \frac{A_{,r}}{A} + \frac{1}{4} \frac{\l(B_{,t}\r)^2}{AB} - \frac{B_{,tt}}{A} + \frac{1}{2} \frac{A_{,t} B_{,t}}{A^2}\r] &=& \l(\Phi_{,r}\r)^2 + \frac{B}{A} \l(\Phi_{,t}\r)^2 - 2B  U(\Phi) \\
\label{e:gtr}
 \frac{B_{,t} B_{,r}}{B^2} + \frac{1}{2} \frac{B_{,t} A_{,r}}{AB} - \frac{B_{,tr}}{B} &=& \Phi_{,t} \Phi_{,r} \\
\label{e:gtettet}
\frac{A_{,rr}}{A} - \frac{A_{,r}}{Ar} - \frac{1}{2} \l(\frac{A_{,r}}{A}\r)^2 - \frac{A_{,r}B_{,r}}{AB}
+ \frac{B_{,rr}}{B} - \frac{B_{,r}}{rB} - \frac{3}{2} \l(\frac{B_{,r}}{B}\r)^2 &=& - 2 \l(\Phi_{,r}\r)^2 \\
\label{e:divt}
\frac{\Phi_{,rr}}{B} - \frac{\Phi_{,tt}}{A} + 2 \frac{\Phi_{,r}}{rB} + \frac{1}{2} \frac{\Phi_{,r} A_{,r}}{AB} + \frac{1}{2} \frac{B_{,r} \Phi_{,r}}{B^2} - \frac{3}{2} \frac{\Phi_{,t}B_{,t}}{AB} + \frac{1}{2} \frac{A_{,t}\Phi_{,t}}{A^2} - U'(\Phi) &=& 0 \ .
\eea
Equations (\ref{e:gtt}, \ref{e:grr}, \ref{e:gtr}, \ref{e:gtettet}) correspond respectively to the components $2E_{tt}$, $2E_{rr}$, $E_{tr}$ and $2\l(E_{\theta\theta}/r^2 - E_{rr}\r)$ of Einstein's equations. Equation (\ref{e:divt}) is the conservation of the energy-momentum tensor.

In the following, one will consider the simplest choice for the potential, $U = m\Phi^2/2$, a free massive Klein-Gordon field, and by suitable scaling we set $m=1$.

\subsection{Mode decomposition}
\label{ss:mode}

As in Ref.\ \cite{SeideS91}, one wishes to study the possibility that there exist periodic, or quasi-periodic solutions, to the system
(\ref{e:gtt}-\ref{e:divt}). Following \cite{SeideS91, FodorFM10} one assumes that $A$ and $B$ contain only even harmonics with respect to time and $\Phi$ only odd ones, which is valid for any symmetric potential $U$. This gives the following mode-decompositions :
\bea
\label{e:modeA}
A \l(r,t\r) &=& 1 + \sum_{j=0}^\infty A_{2j}\l(r\r) \cos\l(\l(2j\r) \omega t\r) \\
\label{e:modeB}
B \l(r,t\r) &=& 1 + \sum_{j=0}^\infty B_{2j}\l(r\r) \cos\l(\l(2j\r) \omega t\r) \\
\label{e:modeP}
\Phi \l(r,t\r) &=& \phantom{1+} \sum_{j=0}^\infty \Phi_{2j+1}\l(r\r) \cos\l(\l(2j+1\r) \omega t\r) \ ,
\eea
where $\omega$ is the frequency of the solution.

\subsection{Asymptotic behavior}
\label{ss:asymptot}
The asymptotic behavior of time periodic solutions of the EKG system is complicated by the fact that the existence of standing wave tails is not compatible with asymptotic flatness.
Heuristically, no matter how small the amplitude of the oscillating tail, due to its slow spatial decay the total mass is infinite. Fortunately one can sidestep this somewhat complicated issue, since there is an intermediate asymptotic region, which is defined by being sufficiently far from the core region of the oscillaton where the first Fourier component of the oscillating tail still dominates, but the mass in this tail is negligible
with respect to the mass in the core.
In Ref.\ \cite{FodorFGR06}, we succeeded to determine numerically the amplitude of the standing wave tail of oscillons in the case of a single scalar field with a self-interaction potential in flat space-time. In this context the amplitude of the oscillons was successfully isolated using a formalism based in the homogeneous solutions of the various operators. The aim of this work is to use similar techniques in the case of the oscillaton. The first step is to assume that there is an intermediate region where space-time can be
considered asymptotically flat. This translates to the following behavior for the metric fields, at large radius :
\bea
\label{e:asymptotA}
A &=& 1 - \frac{r_A}{r} \\
\label{e:asymptotB}
B &=& 1 + \frac{r_B}{r} \ .
\eea
For a solution of the full system, one must have $r_A = r_B\equiv r_0$. This is however not enforced directly in the numerical solution but rather used as a measure of the accuracy of the code.

The situation of the scalar field is more complicated and one has to study in some detail the wave-equation (\ref{e:divt}). If one keeps only the dominating terms (i.e. if one sets $A=1$ and $B=1$), Eq. (\ref{e:divt}) reduces to :
\begin{equation}
\Phi_{,rr} +\frac{2}{r}\Phi_{,r}-\Phi_{,tt} - \Phi = 0 \ .
\end{equation}
$\Phi$ being a sum of odd cosines, the equation for the harmonic $\Phi_n$ is :
\be
\left(r\Phi_{n}\right)_{,rr} + \l(n^2\omega^2-1\r) r \Phi_n = 0 \ .
\ee
Given that $\omega<1$, one has two different cases :
\begin{itemize}
\item For $n=1$, the solution that vanishes at infinity is
\be
\label{e:shphi1}
\Phi_1\l(r\r) = C_1 \displaystyle\frac{\exp\l(-\varepsilon r\r)}{r} \ ,
\ee
with $\varepsilon = \sqrt{1-\omega^2}$.
\item For $n>1$, the solution is oscillatory of the form
\be
\label{e:shphin}
\Phi_n\l(r\r) = C_n \displaystyle\frac{\cos\l(\lambda_n r + \alpha_n\r)}{r} \ ,
\ee
with  $\lambda_n = \sqrt{n^2\omega^2 - 1}$.
\end{itemize}

At this order of approximation, the background is flat Minkowskian, and the phase $\alpha_n$ is constant. However, considering waves on a Schwarzschild background, the phase will have a slow radial dependence. Hence, it will prove useful to get the next order of approximation for $n>1$. In order to do so, we consider the phases $\alpha_n$ as slowly varying
functions of the radius $r$, so that $\alpha_{n,r} \ll \lambda_n$. More precisely, one assumes that $ \alpha_{n,r}$ is of order $\lambda_n / r$. Keeping the first two orders in terms of $1/r$, one gets :
\bea
\label{Pder}
\Phi_{n,r}\l(r\r) &=& -C_n\l(\lambda_n+\alpha_{n,r}\r) \frac{\sin\l(\lambda_n r+\alpha_n\r)}{r} - C_n\frac{\cos\l(\lambda_n r + \alpha_n\r)}{r^2} \\
\label{Pdder}
\Phi_{n,rr}\l(r\r) &=& -C_n\l(\lambda_n^2+2\lambda_n\alpha_{n,r}\r) \frac{\cos\l(\lambda_n r+\alpha_n\r)}{r} + 2C_n \lambda_n \frac{\sin\l(\lambda_n r+\alpha_n\r)}{r^2} \ .
\eea
So it appears that the terms involving $\alpha_{n,r}$ are of the same order as the corrections induced by the $r_0/r$ parts of the metric fields, the equation for the scalar field, up to $1/r^2$ terms, being
\be
\label{e:wave}
\l(1-\frac{r_0}{r}\r) \Phi_{n,rr} + \frac{2}{r}\Phi_{n,r} + n^2 \omega^2 \l(1+\frac{r_0}{r}\r) \Phi_n  - \Phi_n = 0 \ .
\ee
When inserting Eqs. (\ref{Pder}-\ref{Pdder}) into (\ref{e:wave}), the second order terms lead to an equation for the phase $\alpha_n$ (the first order condition has already been used and gives the value of $\lambda_n$). One finds that
\be
- 2 \lambda_n \alpha_{n,r} + \frac{r_0}{r} \l(\lambda_n^2 + n^2 \omega^2\r) = 0,
\ee
which can be integrated to give :
\be
\label{e:delta}
\alpha_n = \frac{r_0}{2} \l(\frac{2\lambda_n^2+1}{\lambda_n}\r) \log r + \delta_n,
\ee
where $\delta_n$ is a constant. One can check that, as assumed, $\alpha_{n,r}$ is indeed of the order of $\lambda_n / r$, in agreement with the starting hypothesis. A detailed study of series solutions for the Klein-Gordon equation on Schwarzschild spacetime can be found in \cite{Elizalde}.

The oscillatory behavior of Eq. (\ref{e:shphin}) illustrates the fact that, in general, time periodic solutions cannot be truly localized, nor can the corresponding space-time be
asymptotically flat. Indeed, if $\Phi_n$ behaves like (\ref{e:shphin}), the terms involving $\Phi$ in Eqs. (\ref{e:gtt}-\ref{e:gtettet}) will decrease only like $1/r^2$, thus being inconsistent with the dominating behaviors of $A$ and $B$ assumed from the start and given by Eqs. (\ref{e:asymptotA}-\ref{e:asymptotB}). It is only due to the smallness of the amplitude of the tail that an ``intermediate asymptotic'' region exists.

In some very particular situations, it can occur that the oscillatory tail is absent thus truly localized time periodic solutions do exist. A famous example is the breather solution in sine-Gordon theory in $1$ dimension. Results from \cite{SeideS91, SeideS94, FodorFM10} strongly suggest that, even if the oscillatory tail is present, its amplitude should be relatively small, so that there exists a region where its influence on the various fields is also small.

\section{Numerical methods}
\label{s:numerical}

\subsection{Spectral expansion}
\label{ss:spectral}

Solutions of the system are sought by making use of the spectral library Kadath \cite{kadath, Grand10}. The setting uses a two-dimensional space, with respect to the coordinates $\l(t, r\r)$. For the time-coordinate, a single domain is used. The physical time $t$ relates to the numerical one $t^\star$ by $t^\star = \omega t$. A spectral expansion is then performed with respect to $t^\star$, using only even cosines for $A$ and $B$ and only odd ones for $\Phi$, in accordance with the mode decompositions (\ref{e:modeA}-\ref{e:modeP}).

For the radial coordinate a multi-domain decomposition is used, similar to the one described in Sec. (2.2) of \cite{Grand10}. Typically one considers a nucleus that contains the origin and several spherical shells that are bounded by two finite radii. However, given the appearance of oscillatory solutions (see Sec. \ref{ss:asymptot}), no compactification of space is used and the equations  are solved only up to a given radius $R_{\rm max}$ at which an appropriate matching is performed (see Sec. \ref{ss:matching}). In each domain the physical radius $r$ is related to the numerical one $r^{\star}$ by an affine-law. In the nucleus one uses $r = R_{\rm nuc} \times r^\star$ with $r^\star \in\l[0,1\r]$ and where $R_{\rm nuc}$ is the radius of the nucleus. In the shells one uses $r = \l(\displaystyle\frac{R_{\rm outer}-R_{\rm inner}}{2}\r) r^\star + \l(\displaystyle\frac{R_{\rm outer}+R_{\rm inner}}{2}\r)$, with $r^\star \in \l[-1, 1\r]$ and where $R_{\rm inner}$ and $R_{\rm outer}$ are the inner and outer radii of the domain. Spectral expansion is performed with respect to $r^{\star}$. In the nucleus, and to account for the fact that the fields are even near the origin, only even Chebyshev polynomials are used (this is also the reason for the different range of variation of $r^\star$). Spectral expansion is performed with respect to standard Chebyshev polynomials in the various shells.

For instance, in a given shell, $A$ is approximated by $A \approx \displaystyle\sum_{j=1}^{N_t}
\displaystyle\sum_{i=0}^{N_r} A_{ij} \cos\l(2j t^\star\r) T_i\l(r^{\star}\r)$. $N_t$ and $N_r$ are the number of coefficients  with respect to $t^{\star}$ and $r^{\star}$. $T_i$ denotes the $i^{\rm th}$ Chebyshev polynomial. The $A_{ij}$ are the spectral coefficients of $A$.

Let us finally point out that all the divisions by $r$ that appear in the equations concern quantities that are odd near the origin (like $A_{,r}$). Such ratios are then easily computed using the spectral expansion because the ratio of $T_{2i+1} / r$ can be exactly expressed as a sum of $T_{2i}$. The division is said to take place in the coefficient space.

\subsection{Matching criteria at the outer boundary}
\label{ss:matching}

Given that the computational domain can not be extended to infinity, one needs to derive appropriate outer boundary conditions for the various fields. They are based on the asymptotic behaviors in the ``intermediate region'' found in Sec. \ref{ss:asymptot}. The idea is to match the various fields to the solutions of the dominating operators, at large radius. More explicitly, let us consider a purely radial function $f\l(r\r)$ that must be matched to the solution $g\l(r\r)$ of the operator. By hypothesis, one demands that, at a large matching radius $R$, $f$ is close to a solution of the form $C g\l(r\r)$, where $C$ is a constant not known {\em a priori}. The continuity of $f$ and its radial derivative $f_{,r}$ at the matching radius gives the following conditions:

\bea
\label{e:matchf}
f\l(R_{\rm max}\r) &=& C g\l(R_{\rm max}\r) \\
\label{e:matchdf}
f_{,r} \l(R_{\rm max}\r) &=& C g_{,r}\l(R_{\rm max}\r).
\eea

The constant $C$ can be eliminated to get a boundary condition for $f$ that involves only the function $g$ :

\be
\label{e:bc}
\l[f g_{,r} - f_{,r} g \r] \l(R\r) = 0.
\ee
Once the function $f$ is known, the constant $C$ can be recovered by making use of either Eq. (\ref{e:matchf}) or (\ref{e:matchdf}).

This matching technique is used to match $A$ (resp. $B$) to a solutions of the type $1 - \displaystyle\frac{r_A}{r}$ (resp. $1-\displaystyle\frac{r_B}{r}$), where the constant $r_A$ (resp. $r_B$) plays the role of $C$ in Eqs. (\ref{e:matchf}) and (\ref{e:matchdf}). Let us note that for those two fields, one only matches the harmonics which are time-independent, the other ones being simply set to zero.

For the scalar field, the matching functions depend on the order $n$ of the harmonic considered and are given by Eq. (\ref{e:shphi1}) for $n=1$ and Eq. (\ref{e:shphin}) otherwise. In the expression (\ref{e:delta}) of the slowly varying phase, one will use $r_B$ for the value $r_0$, as both $r_A$ and $r_B$ are expected to converge to $r_0$ as the matching radius $R_{\rm max}$ increases.

\subsection{Minimization of the oscillatory tail}
\label{ss:minimization}

In the asymptotics given in Sec. \ref{ss:asymptot}, for each $n>1$, there is a constant phase $\delta_n$, that can be chosen freely (see Eq. (\ref{e:delta})). In practice, one wishes to construct solutions that are as close as possible to truly localized ones. Therefore it is desirable to make the oscillatory tails as small as possible. As seen in Sec. \ref{ss:asymptot}, the dominating mode for which the oscillations appear is $\Phi_3$ therefore we try to minimize $C_3$ ($C_i$ being the amplitude of the oscillatory tail of the $i^{\rm th}$ mode, as seen in Sec. \ref{ss:matching}).

An additional simplification comes from the fact that the oscillatory tails of the other modes $n>3$ have a very small influence on $C_3$, since their amplitudes, $C_n$, are decreasing very rapidly for increasing values of $n$. In other words, $C_3$ is almost independent of $\delta_n$ for $n>3$. So the minimization of $C_3$ as a function of only one constant phase $\delta_3$ is a very good approximation to the true minimum. This is done using a standard golden section search algorithm (see for instance \cite{numrecipe}). The amplitude, $C_3$, never gets so small that the other $C_n$'s need be considered (see Sec. \ref{ss:results}).

\subsection{The numerical system}
\label{ss:system}

In the context of the spectral methods implemented by the library Kadath, a system of partial differential equations on the fields is transformed into a set of algebraic equations for the unknowns that are the coefficients of the various fields. The non-linear system is solved by making use of a Newton-Raphson iteration: starting from an initial guess that is as good as possible, the solution is found by iteration. At each step, the linearized system (with respect to the unknowns) is inverted (see Sec. 5 of \cite{Grand10} for more details).

The set of equations (\ref{e:gtt}-\ref{e:divt}) is redundant, meaning there are more equations than unknowns. In order to select an appropriate subset of equations let us recall that the actual number of unknowns of a field depends on the exact spectral basis, some coefficients not being true degrees of freedom (see Sec. 3.5 of \cite{Grand10} for a detailed discussion on that). The same is true for the equations that each give a number of algebraic equations that depend on the spectral basis of the result. In order to ensure that the number of unknowns is identical to the number of equations, one then needs to select equations that have the same spectral basis as the fields. Eq (\ref{e:divt}) has the same basis as the scalar field itself, whereas Eq (\ref{e:gtr}) has both a different radial and temporal basis from the ones of $A$ and $B$ (because the equations involve one derivative in each dimension). However the three equations (\ref{e:gtt},\ref{e:grr},\ref{e:gtettet}) are consistent with the basis of the metric fields. There is no argument to discard one more than the other and one just has to check, after solution that the ``forgotten'' equation is indeed verified. It is found that the set of Eqs. (\ref{e:gtt}) (\ref{e:grr}) and (\ref{e:divt}) lead to a numerical solution that fulfills the full set of equations (see Sec. \ref{ss:tests}).

In the context of Kadath, partial differential equations are dealt with by means of a tau-method (see for instance \cite{canuto, living}). In each domain, the equations are solved by demanding that the coefficients of the residual vanish: in a sense, one solves the equations in the coefficient space. Depending on the order of the equations, the conditions corresponding to the last coefficients must be relaxed to enforce continuity of the solution and appropriate boundary conditions. The number of conditions that must be relaxed is known as the order of the method and is closely related to the number of homogeneous solutions of the operators. The functions being periodic, there is no need for any boundary conditions with respect to time.

For the radial coordinate, the situation depends on both the equation and the type of domain. The radial derivative of the highest order in Eq. (\ref{e:gtt}) is $B_{,rr}$ so that one will consider Eq (\ref{e:gtt}) to be associated with the field $B$. $B_{,rr}$ has two homogeneous solutions: $C$ and $r$. However, in the nucleus, the function $r$ is not admissible because, by construction, we restricted ourselves to functions $B$ that are symmetric near the origin. So we have only one admissible homogeneous solution in the nucleus and two in the various shells so that we have to use a first order tau-method in the nucleus and a second order one in the different shells. Both the variable $B$ and its radial derivative $B_{,r}$ must be matched at the interface in the various domains and one must also supply an outer boundary. One can easily check that the number of conditions discarded by the tau method is equal to the number of matching and boundary conditions.

Given that Eq. (\ref{e:gtt}) is associated with the variable $B$, one has to consider Eq. (\ref{e:grr}) as an equation for $A$. The highest order derivative is $A_{,r}$ which admits only one homogeneous solution. Equation (\ref{e:grr}) is then solved by using a first order tau-method in every domain, supplemented by the matching of $A$ at each interface and an appropriate outer boundary condition. The situation for $\Phi$ is similar to the case of $B$. Let us point out that associating variables with given equations may seem a bit dubious, given that the whole system is coupled, but it does provide very useful guideline in constructing a numerical system that is well-posed. The whole situation is summarized in Table \ref{t:system}.

\begin{table}

\centering
\caption[]{\label{t:system}
Construction of the numerical system to be inverted. The first two columns show the formal association of each unknown to an equation. The order of the $tau$-methods in both the nucleus and the shells are then given, as well as the quantities that must be matched at the boundaries between the domains. The last column gives the function to which each harmonic is matched at the outer radius $R_{\rm max}$.
}
\begin{tabular}{| c | c |  c | c  | c | c | }
  \hline
  Variable & Equation & $\tau$-order nucleus & $\tau$-order shells & Matching & Outer matching \\
\hline
$A$ & (\ref{e:grr}) & $1$ & $1$ & $A$ & $1-\displaystyle\frac{r_A}{r}$ for $n=0$ \\
    &               &     &     &     & $0$ otherwise \\

  \hline

$B$ & (\ref{e:gtt}) & $1$ & $2$ & $B$ and $B_{,r}$ & $1+\displaystyle\frac{r_B}{r}$ for $n=0$ \\
    &               &     &     &     & $0$ otherwise \\
\hline
$\Phi$ & (\ref{e:divt}) & $1$ & $2$ & $\Phi$ and $\Phi_{,r}$ & Eq. (\ref{e:shphi1}) for $n=1$ \\
    &               &     &     &     & Eq. (\ref{e:shphin}) for $n>1$ \\
\hline
  \end{tabular}
\end{table}

\section{Numerical results}
\label{s:results}
\subsection{Setting}
\label{ss:settings}

As observed in \cite{FodorFM10}, the solutions are more and more spatially extended as $\omega$ gets closer to $1$, so that it is convenient to work with the variable $\rho = r   \varepsilon$, where $\varepsilon = \sqrt{1-\omega^2}$. In terms of $\rho$ the geometry of the solutions for various $\omega$ is rather similar. Using this fact as a guideline, we chose the radial domains to be, in terms of $r$, $\l[0, 1/\varepsilon \r]$,   $\l[1/\varepsilon , 2/\varepsilon\r]$, $\l[2/\varepsilon , 4/\varepsilon\r]$ for the three first ones and $\l[4 i /\varepsilon , 4 \l(i+1\r)/\varepsilon\r]$ afterwards. The size of the last domains remains fixed to $4/\varepsilon$ because one wishes to be able to resolve the oscillatory tails that appear in $\Phi$ and so one must ensure that the size of the domains does not get too big, with respect to the wavelength of the oscillations. This setting, along with the number of radial coefficients used, seems to be satisfactory, as can be seen in Sec.\ \ref{ss:results}. The outer matching radius can be varied by changing the number of domains $N_d$. As a standard value one uses $N_d=20$, which corresponds to a matching radius of $R_{\rm max}=68/\varepsilon$.

As far as the number of coefficients are concerned, one uses three different resolutions, a low one with $N_r=13$ radial coefficients and $N_t = 5$ coefficients in time, a medium one with $N_r = 17$ and $N_t=7$ and a high resolution with $N_r = 33$ and $N_t=9$. The system is iterated until the residuals reach a threshold of $10^{-8}$ for the low and medium resolutions and of $10^{-10}$ for the high resolution. The setting of the threshold is somewhat empirical and is related to the precision one can expect given a number of coefficients (which is very difficult to assess a priori in the general case). A threshold too big will obviously result in a poor precision, no matter what the number of coefficients is. On the other hand, if the threshold is too small, it can prevent the code from converging, if the number of coefficients is not sufficient. Depending on the cases, the overall precision of the resolution is limited either by the number of coefficients or by the value of the threshold (see the discussion about Fig. \ref{f:errors}).

The constant phases $\delta_n$ are fixed to zero for $n\not= 3$, whereas $\delta_3$ is obtained by minimizing the coefficient $C_3$. The search for the minimum is stopped when a precision of $10^{-2}$ is achieved on the value of $\delta_3$. It will indeed appear that even if $C_3$ varies greatly as a function of $\delta_3$, the curve is very flat near the minimum value so that a greater precision on $\delta_3$ is not required to get a precise value of $C_3$. For instance, for $\omega=0.86$, this is sufficient to get a relative precision on $C_3$ of about $10^{-5}$, which is smaller than other sources of errors.

As an initial guess for starting the Newton-Raphson iteration, one uses the first order expansion in terms of $\varepsilon$ given in Sec. III-B of \cite{FodorFM10}, for $\varepsilon \approx 0.1$. Once some solutions are known for $\omega$ close to $1$, they can be used as an initial guess for computing new configurations, by slowly varying $\omega$.

\subsection{Tests}
\label{ss:tests}

\begin{figure}[htb]
\includegraphics[width=0.5\textwidth]{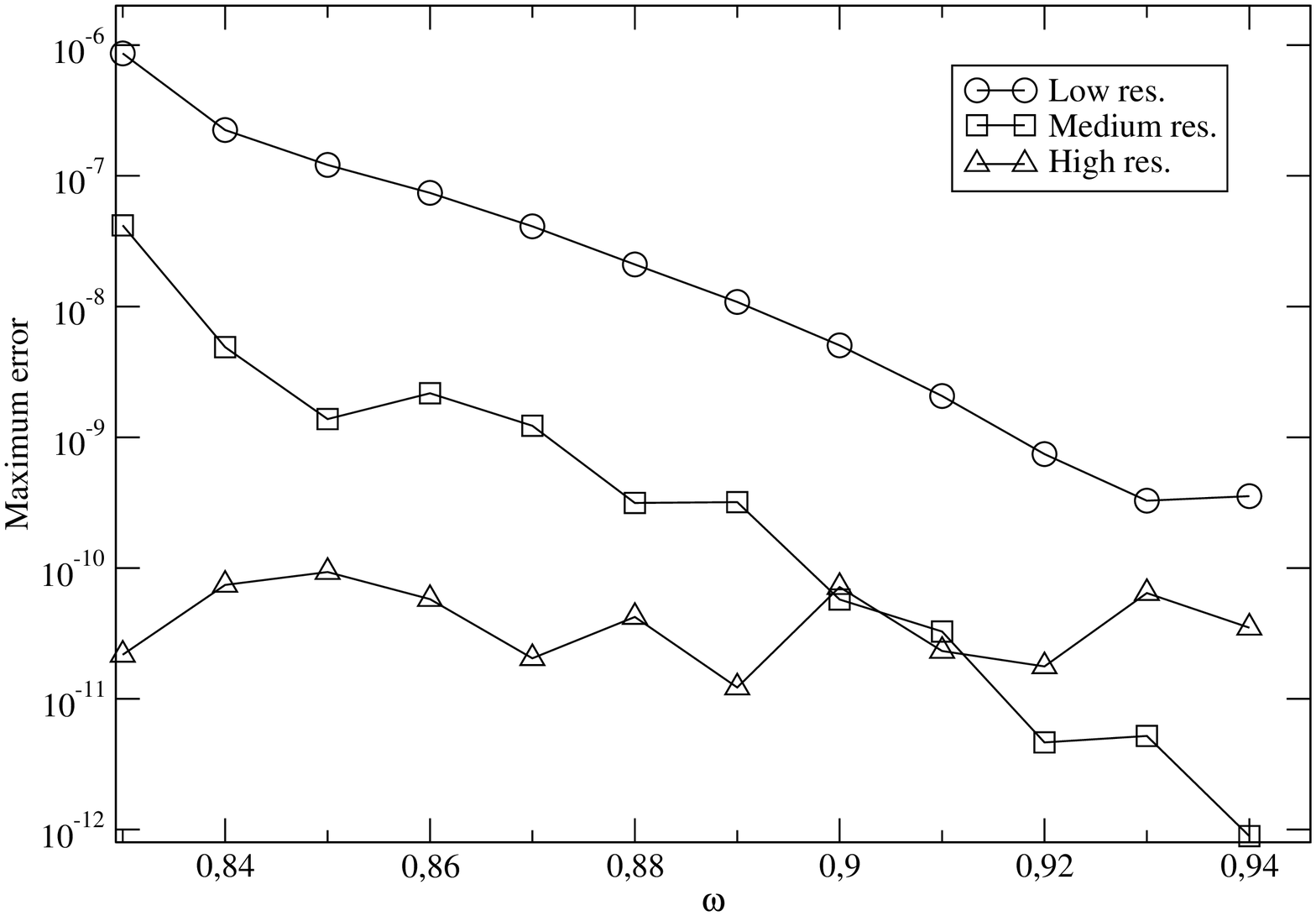}
\caption{ \label{f:errors}
Maximum error on the full set of equations, as a function of $\omega$, for the three different resolutions.}
\end{figure}

In order to assess the precision reached by the numerical code, one can check the residual error on the equations that have not been used explicitly in finding the solution, i.e. Eqs. (\ref{e:gtr}) and (\ref{e:gtettet}). The maximal error measured by the highest coefficient of the residuals in shown in Fig. \ref{f:errors}, for the three different resolutions and various values of $\omega$. In general the lower $\omega$, the bigger the various modes are so that one needs more coefficients to achieve a given precision (this is similar to the case of the Kerr black hole ; see Fig. 9 of \cite{Grand10}). So, for a fixed number of coefficients the expected errors coming from the spectral representation should be higher for lower values of $\omega$. This is what is observed for the lowest resolution until an error of around $10^{-9}$ is achieved. This saturation is coming from the value of the threshold set to stop the Newton-Raphson iteration. The case of the highest resolution is also easy to understand. The fact that the curve is almost flat at a level of $10^{-11}$ illustrates the fact that, in this case, the factor limiting the precision is not the number of coefficients but solely the value of the threshold, set in this case to $10^{-10}$. The case of the medium resolution can seem a bit puzzling. As for the low resolution, the error decreases when $\omega$ increases, showing that one is dominated by the errors coming from the spectral approximation. However, contrary to what could be expected, the error does not show any sign of saturation at the level of the threshold of the Newton-Raphson iteration. The error is even getting smaller than in the high resolution case, for high values of $\omega$. The reason for that is to be found in the precise way the Newton-Raphson iteration proceeds. The iteration is stopped as soon as the error gets below the threshold. In general when this happens, the error is just below the threshold, and so of the same order of magnitude (this is the case for the highest resolution for instance). However, in the medium resolution, the first value below the threshold is actually much lower that the threshold itself. The Newton-Raphson iteration would have stopped at the same iteration even for a much smaller threshold. This is why the curve for the medium resolution does not show signs of saturation. This is nothing profound and is only coincidental for this particular problem.

We could have tried to reduce the residual error for the highest resolution by lowering the threshold but this could lead to convergence issues coming from round-off errors, the code using only double precision. Moreover, the precision reached is sufficient for our purpose, which is the precise extraction of the oscillatory tails (see Sec. \ref{ss:results}). It is worth mentioning that the errors shown in Fig. \ref{f:errors} do not measure the accuracy at which the oscillatory tails can be extracted. It will indeed appear clearly in Sec. \ref{ss:results} that the results from the high resolution cases are more accurate than those from the medium one, especially when the phase $\delta_3$ is concerned. Let us finally mention that using Eq. (\ref{e:gtettet}) in place of Eq. (\ref{e:grr}) leads to the appearance of spurious solutions that do not fulfill the full set of equations.

\begin{figure}[htb]
\includegraphics[width=0.5\textwidth]{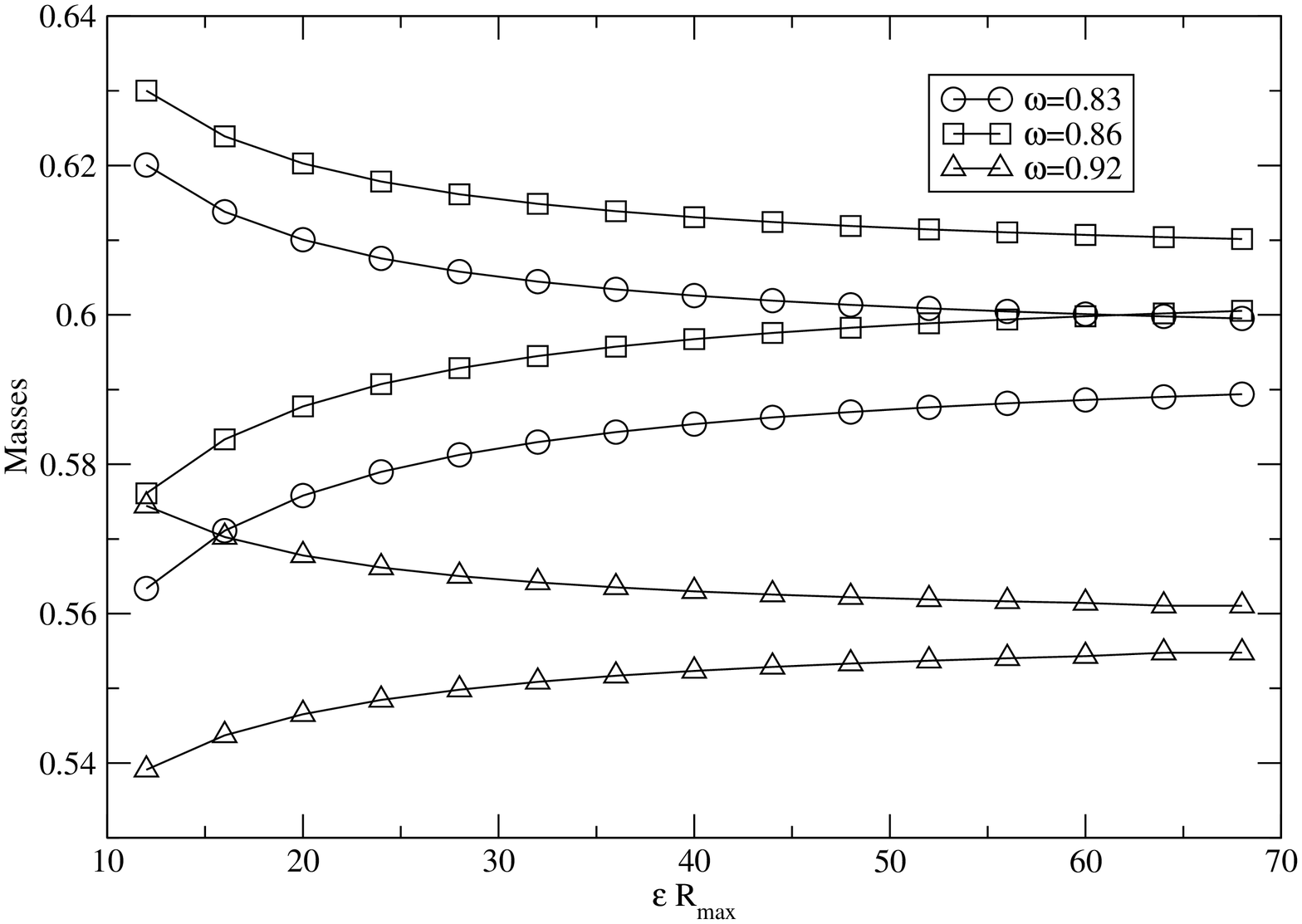}
\includegraphics[width=0.5\textwidth]{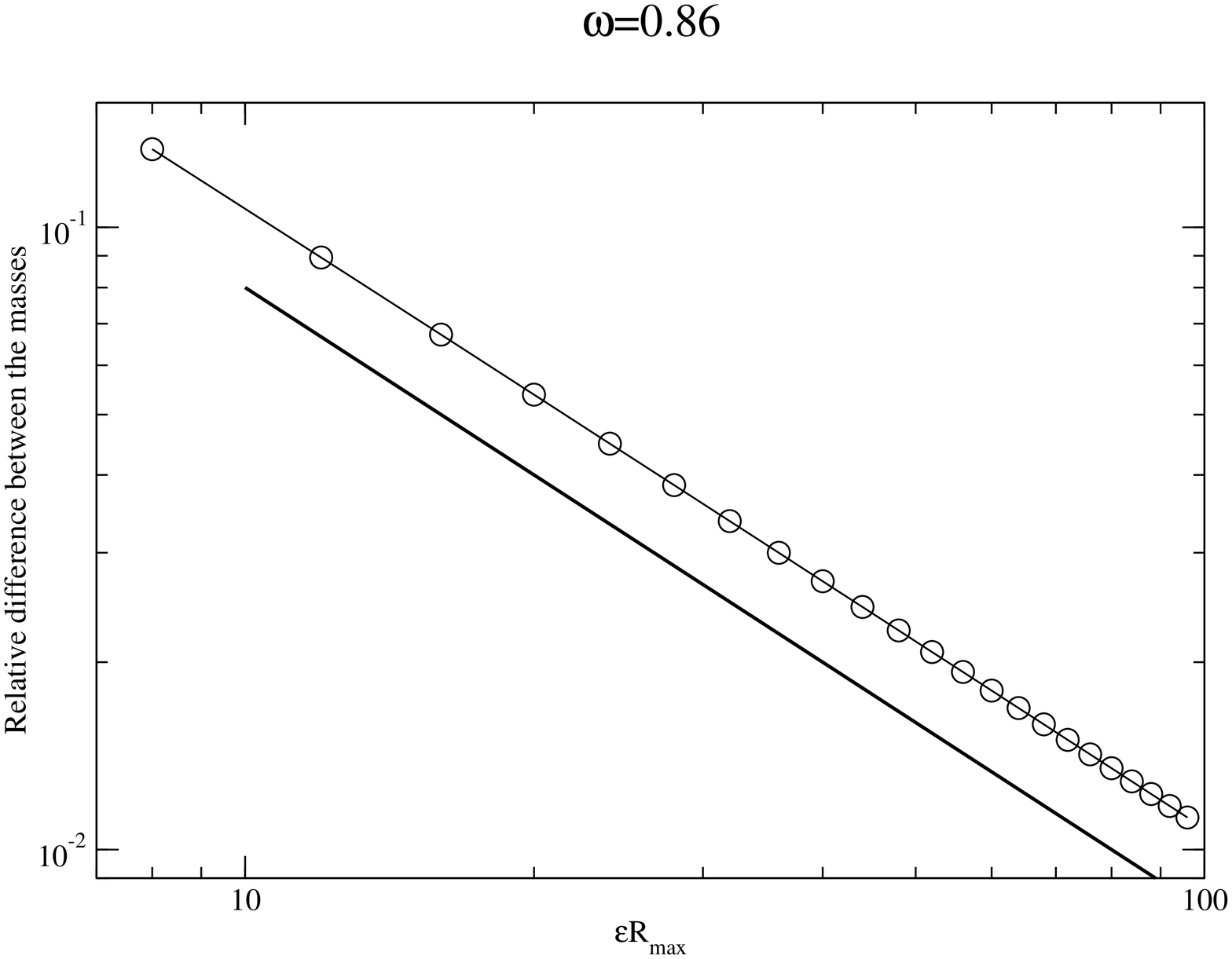}
\includegraphics[width=0.5\textwidth]{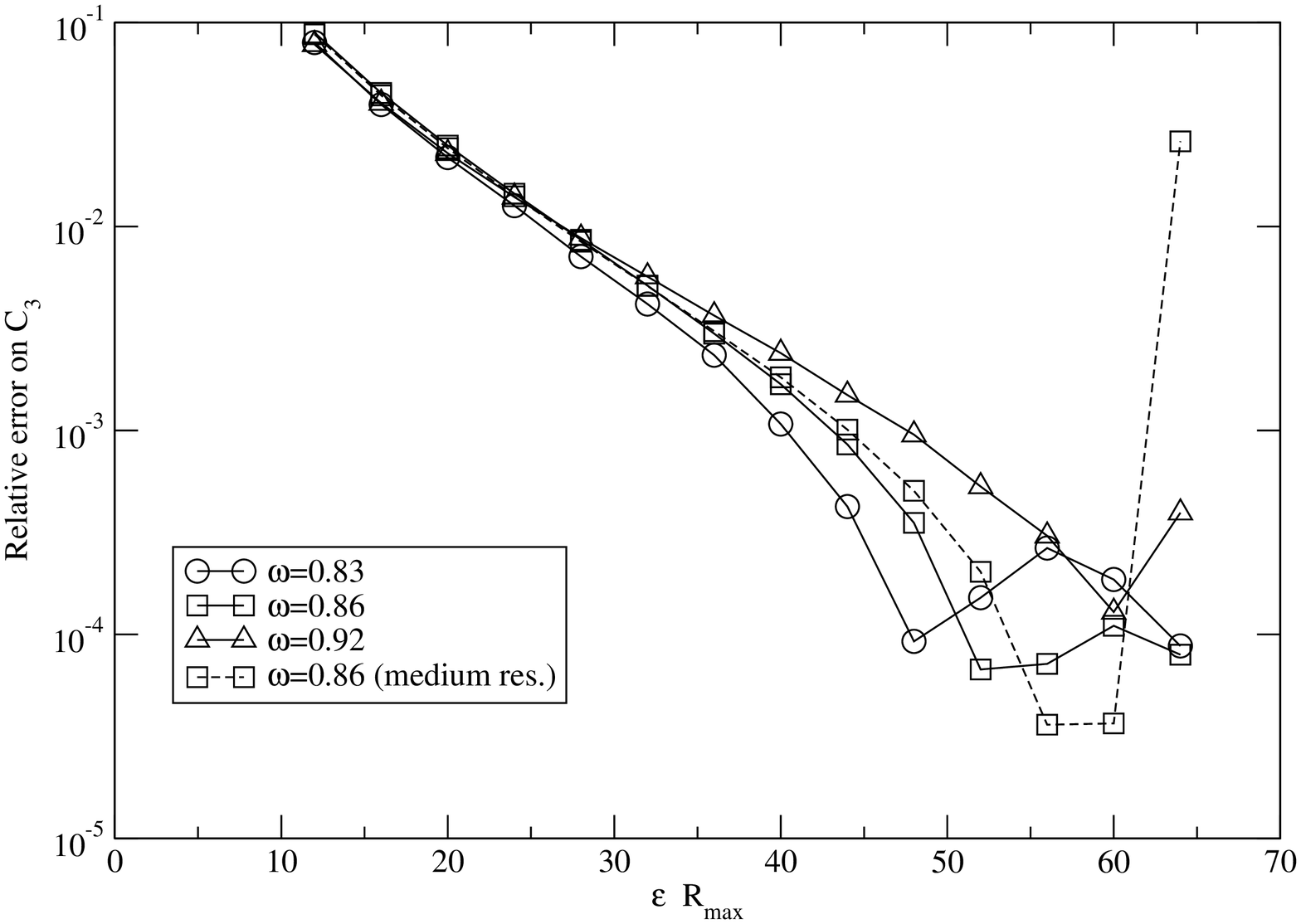}
\caption{ \label{f:conv}
The first panel shows the behavior of the masses as a function of $R_{\rm max}$. The increasing (resp. decreasing) curves correspond to the masses extracted from $A$ (resp. $B$). The circles correspond to $\omega=0.83$, the squares to $\omega=0.86$ and the triangles to $\omega = 0.92$. The second panel shows the relative difference between $M_A$ and $M_B$, as a function of the radius, for $\omega=0.86$. The curve decreases as $1/r$ as can be seen by comparing it with the bold curve, which is exactly $1/r$. The third panel shows the convergence of $C_3$ as a function of $R_{\rm max}$ for three different values of $\omega$. In the case of $\omega=0.86$ we also show the results for the medium resolution (squares and dashed curve).
}
\end{figure}

Figure \ref{f:conv} shows how the numerical results depend on the matching radius $R_{\rm max}$. It is a very important test that validates the matching techniques explained in Secs. \ref{ss:asymptot} and \ref{ss:matching}. In the first panel, the masses are shown as a function of $R_{\rm max}$. In three spatial dimensions, the masses can be read from either $A$ or $B$ and relate to the constants $r_A$ and $r_B$ by $M_A = r_A/2$ and $M_B = r_B/2$ (see Sec. \ref{ss:asymptot} for the definition of $r_A$ and $r_B$). As expected, $M_A$ and $M_B$ converge to the same value as $R_{\rm max}$ increases, $M_A$ being an increasing function of $R_{\rm max}$ whereas $M_B$ is a decreasing one. The convergence is clearly seen in the second panel, where the relative difference between $M_A$ and $M_B$ is plotted, for $\omega=0.86$. The curve goes to zero almost exactly like $1/r$, as expected. The third panel shows the relative error in the value of $C_3$ as a function of $R_{\rm max}$ for three different values of $\omega$. More precisely, we show the relative difference with the value obtained for our largest $R_{\max}$ which is $68 / \epsilon$. For the three curves the minimization with respect to $\delta_3$ is not performed in each case but $\delta_3$ is fixed to the value found for $R_{\max} = 68 / \epsilon$. The matching procedure is validated as the relative difference decreases fast as $R_{\rm max}$ increases. There is a saturation of the relative error at a level of about  $10^{-4}$. The fact that the saturation depends on the resolution (as shown by the dashed curved in the case of $\omega=0.86$) implies that this is an effect of the overall precision of the code ; one is indeed talking of a relative difference of $10^{-4}$ on a quantity which is already very small ($10^{-5}$ for $\omega=0.83$ and $10^{-10}$ for $\omega=0.92$). Those curves indicate that one can expect to reach a relative precision of around $10^{-4}$ on the value of $C_3$.

\begin{figure}[htb]
\includegraphics[width=.5\textwidth]{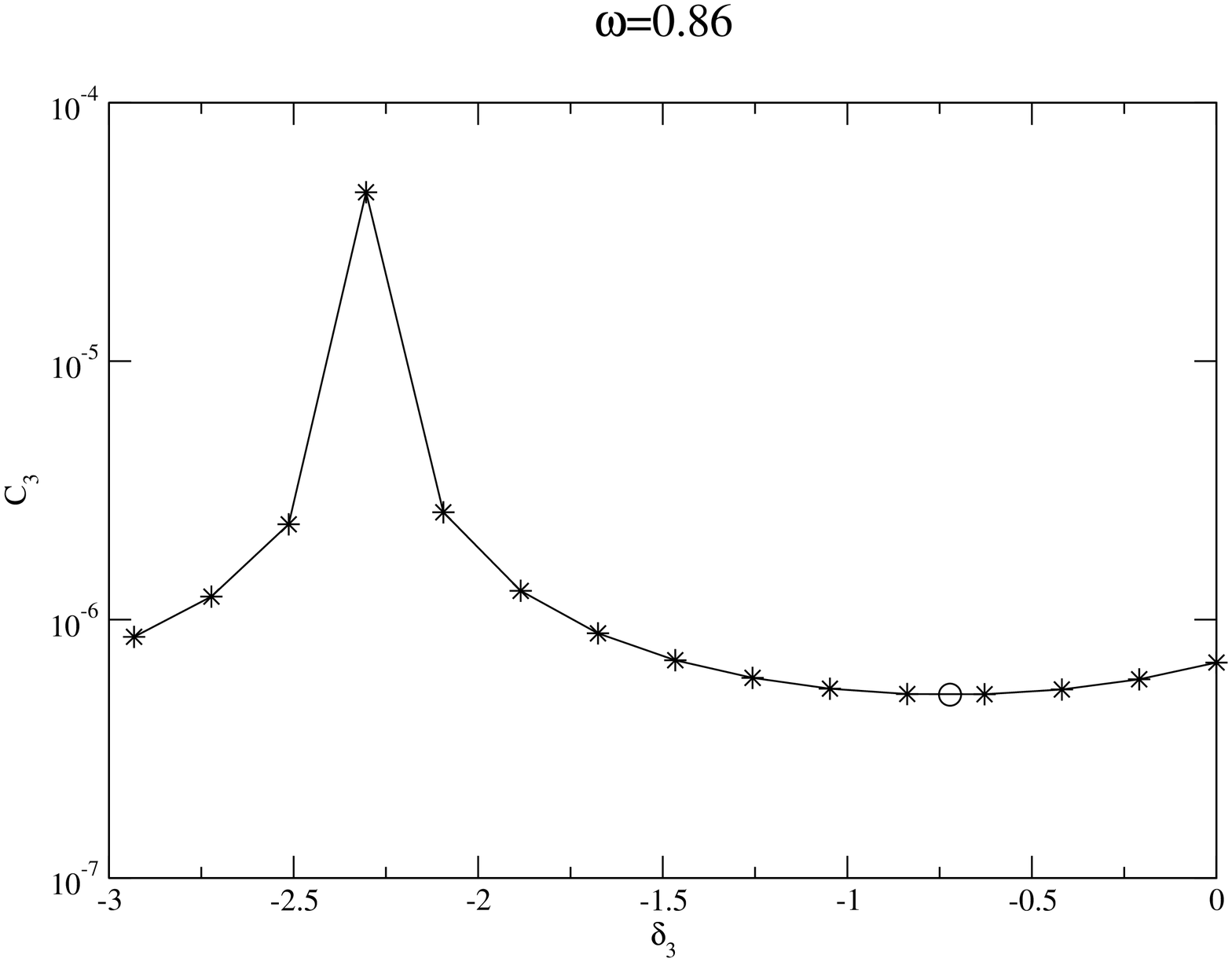}
\includegraphics[width=.5\textwidth]{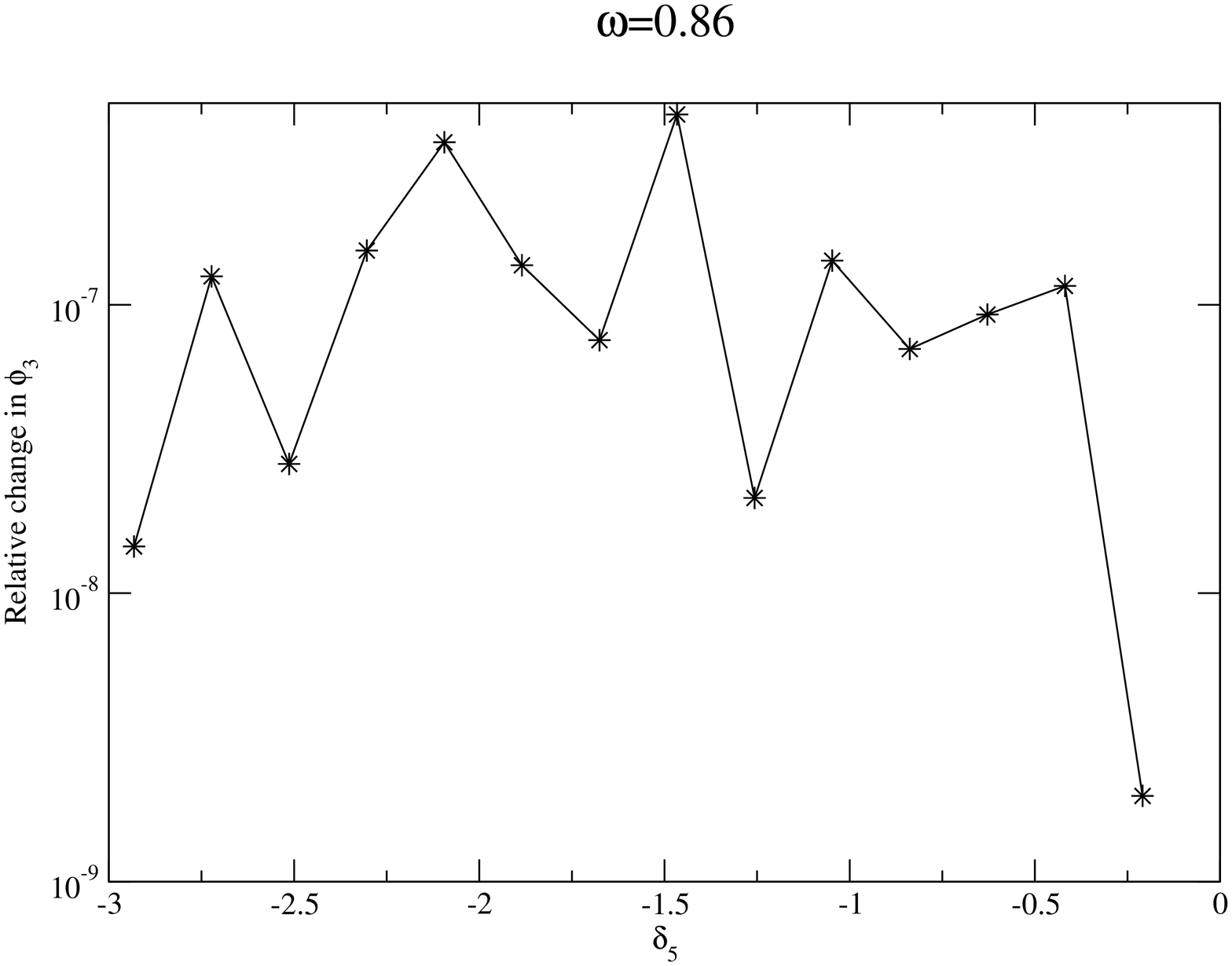}
\caption{ \label{f:deffect}
On the first panel, for $\omega=0.86$, the value of $C_3$ is given as a function of the phase $\delta_3$. The circle corresponds to the position of the minimum found with the golden section search algorithm. The second panel shows the influence on the next constant phase $\delta_5$, where $\delta_3$ is fixed to its value for $\delta_5=0$.}
\end{figure}

Figure \ref{f:deffect} shows the influence of the phases $\delta_n$ on the value of $C_3$, for $\omega=0.86$. The first panel shows $C_3$ as a function of $\delta_3$ and one can see that the value varies by two orders of magnitude. The value of the minimum found by the golden section search algorithm is indicated by the circle. The second panel shows the influence of $\delta_5$ on $C_3$. More precisely one shows the relative difference $\l|C_3\l(\delta_5\r) - C_3\l(\delta_5 = 0\r)\r|/C_3\l(\delta_5 = 0\r)$. $\delta_3$ is fixed to the value found by minimization with $\delta_5=0$. The other phases are fixed to zero. As expected it is very small, the variations on the value of $C_3$ being of the order of $10^{-7}$. This validates the fact that one seeks the minimum of $C_3$ by varying only the phase $\delta_3$.

\subsection{Results}
\label{ss:results}

\begin{figure}[htb]
\includegraphics[width=.5\textwidth]{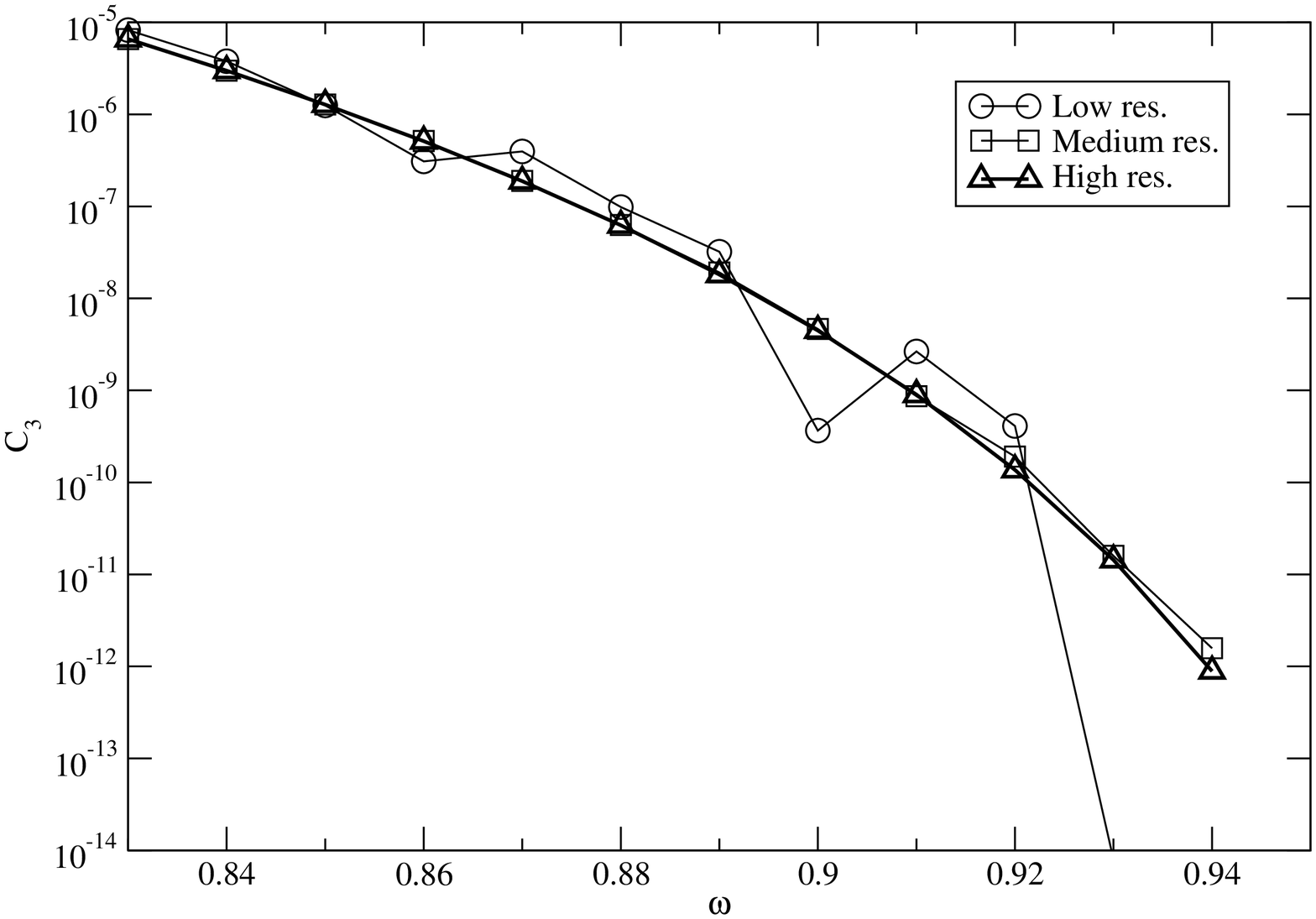}
\includegraphics[width=.5\textwidth]{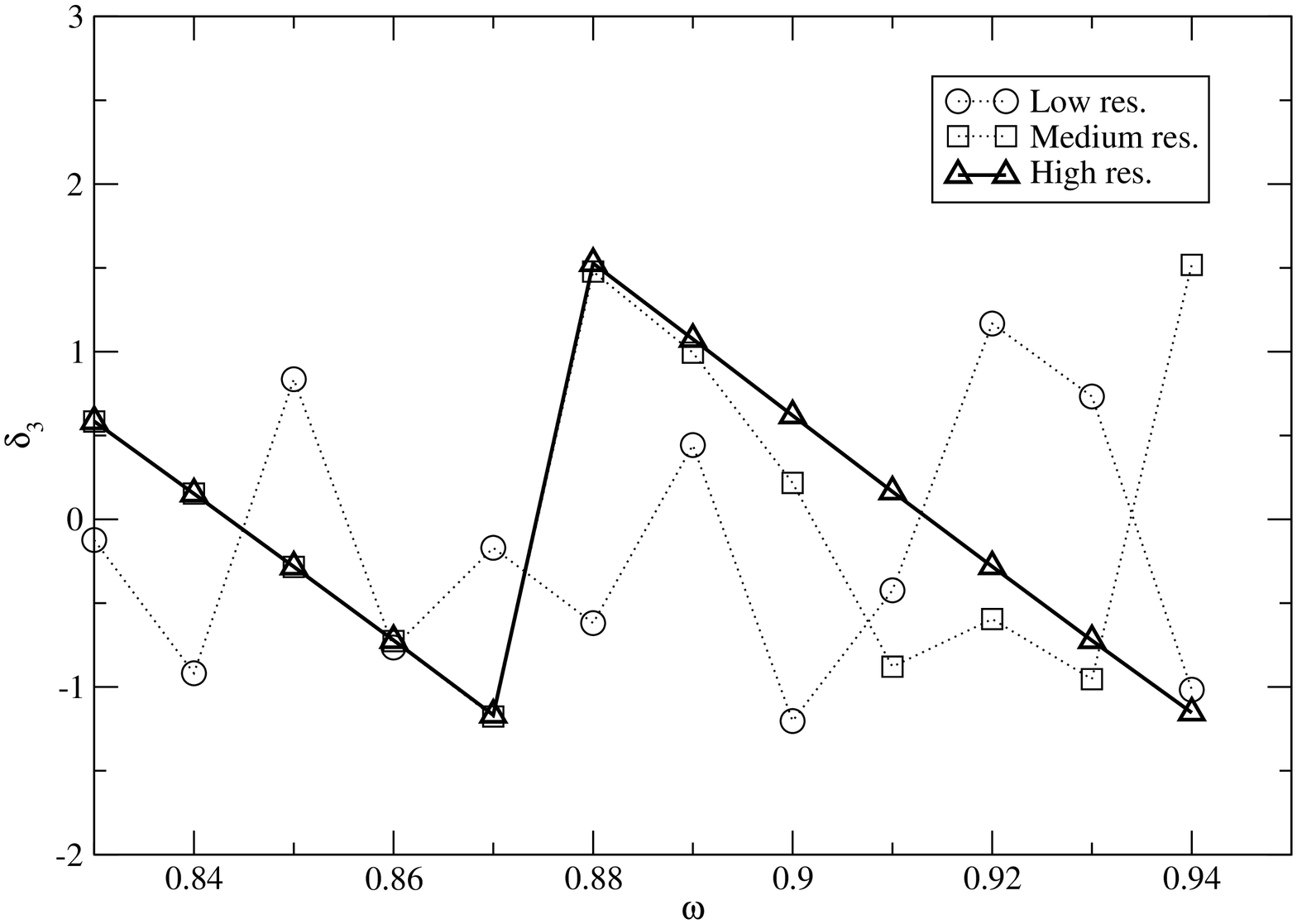}
\caption{ \label{f:c3}
The first panel shows the minimum value $C_3$, as a function of $\omega$, for the three different resolutions. For the same configurations, the second panel shows the phase $\delta_3$ for which the minimum is attained.}
\end{figure}

The main result of our simulations is to show that there is no value of $\omega$ for which the oscillatory tail vanishes. This can be seen on the first panel of Fig. \ref{f:c3}, where the minimum value of $C_3$ is plotted, as a function of $\omega$, for the three different resolutions. The fact that $C_3$ does not vanish implies that there exist no truly periodic solutions of the system. Nevertheless, $C_3$ is very small and thus one can have very long-lived solutions as observed in \cite{SeideS91, SeideS94} for instance. The constant phase $\delta_3$ for which the minimum value of $C_3$ is attained is shown in the second panel of Fig. \ref{f:c3}. It appears that the low resolution is not precise enough to give a good location of the minimum and that the medium one seems to be accurate only for low values of $\omega$ for which the various modes $\Phi_n$ are bigger. The results from the highest resolution solution are very smooth and one can expect them to be accurate on the whole range of $\omega$. Let us mention that even if the phase is not always very precisely determined, the value of $C_3$ is obtained with a relatively good accuracy, the curve being rather flat near the minimum (see the first panel of Fig. \ref{f:deffect}).

\begin{figure}[htbp]
\includegraphics[width=.5\textwidth]{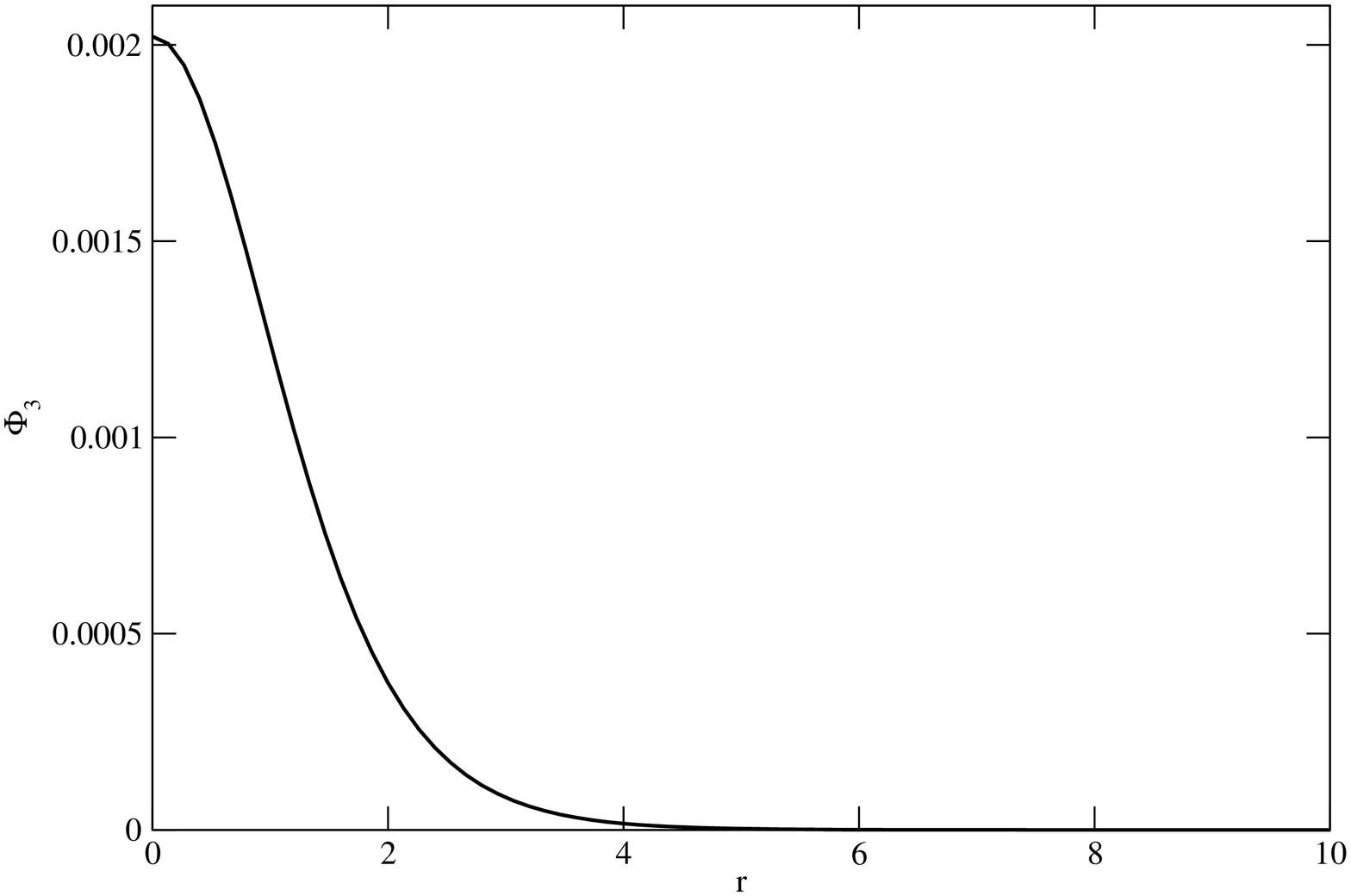}
\includegraphics[width=.5\textwidth]{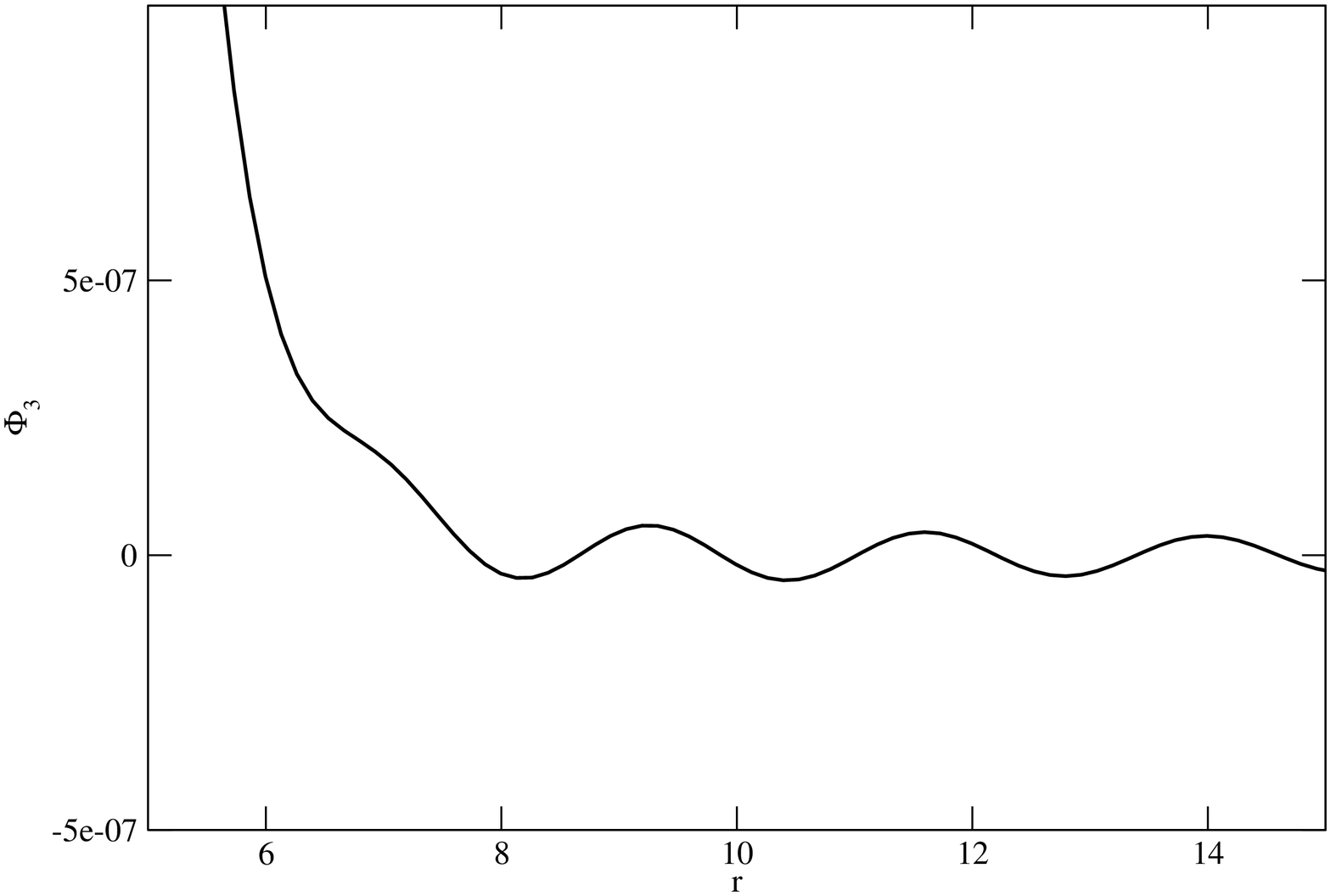}
\includegraphics[width=.5\textwidth]{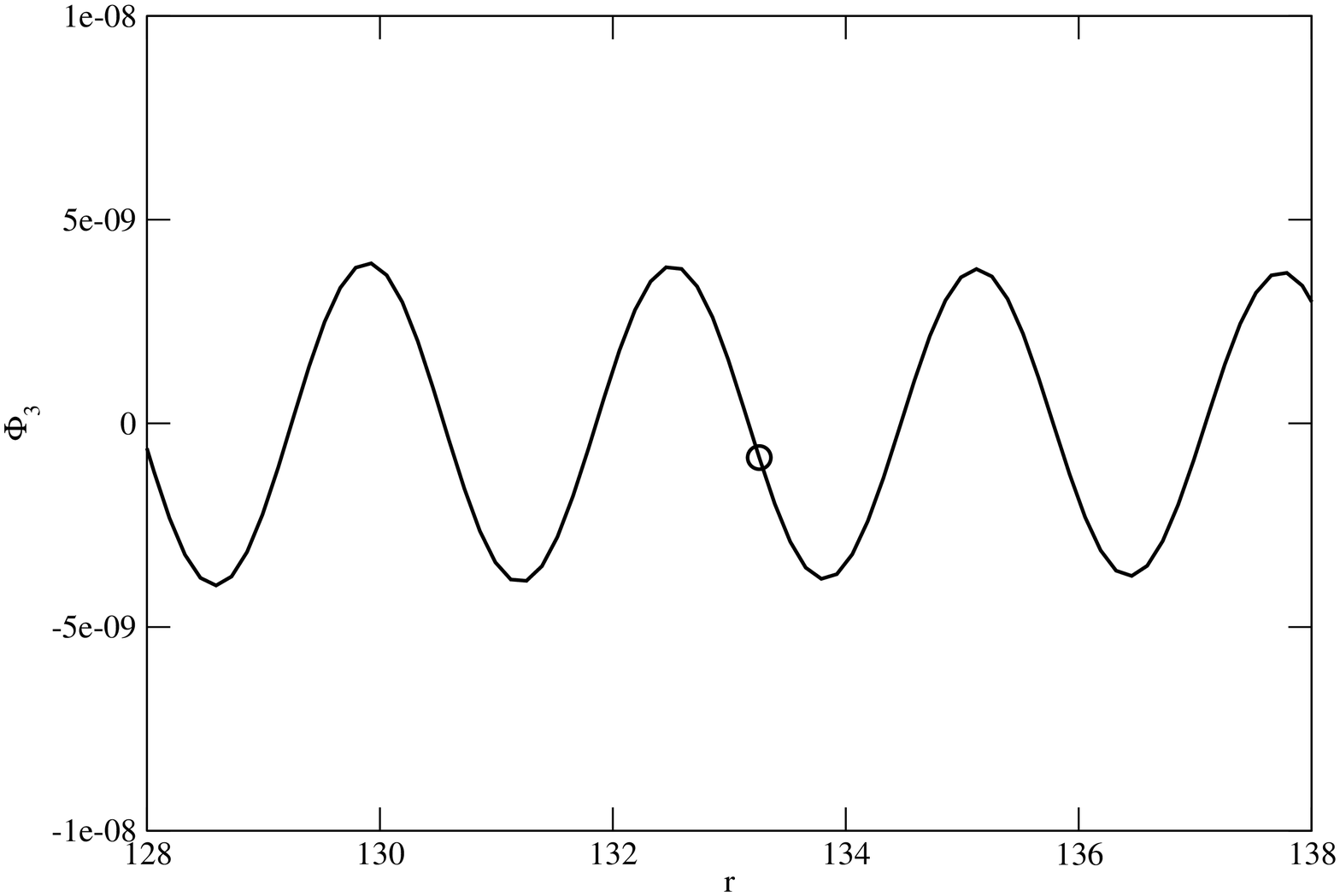}
\caption{ \label{f:phi3}
Value of $\Phi_3$ as a function of $r$, for $\omega=0.86$. The first panel shows the region near the origin, the second one the region where the oscillations begin to dominate and the third one the region around the outer matching radius $R_{\rm max}$. The position of $R_{\rm max}$ is denoted by the circle.}
\end{figure}

As an illustration, in Fig. \ref{f:phi3}, we show the mode $\Phi_3$ as a function of $r$, for $\omega=0.86$. This mode is the first that is matched to an oscillatory solution. The different panels show the field in different regions: i) near the origin in the first panel, ii) in transition region where the oscillations begin to dominate in the second panel and iii) the region close to the matching radius $R_{\rm max}$ in the third panel. In this latter case, the circle denotes the value of $R_{\rm max}$. For $r>R_{\rm max}$ there is no numerical solution so we plot the analytical formula (\ref{e:shphin}) instead. The numerical solution and its analytical continuation are indistinguishable by eye.

\begin{figure}[htbp]
\includegraphics[width=.5\textwidth]{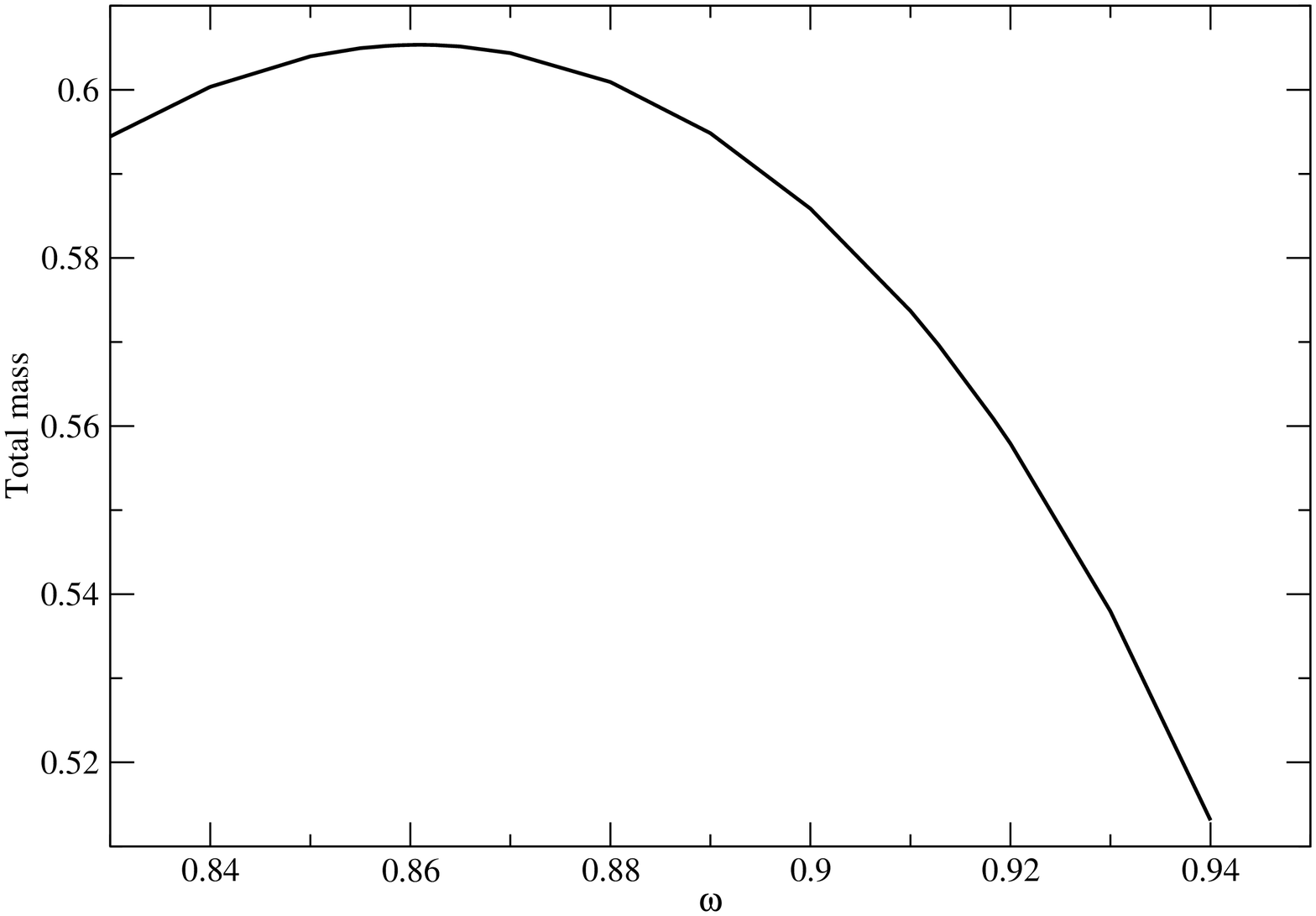}
\includegraphics[width=.5\textwidth]{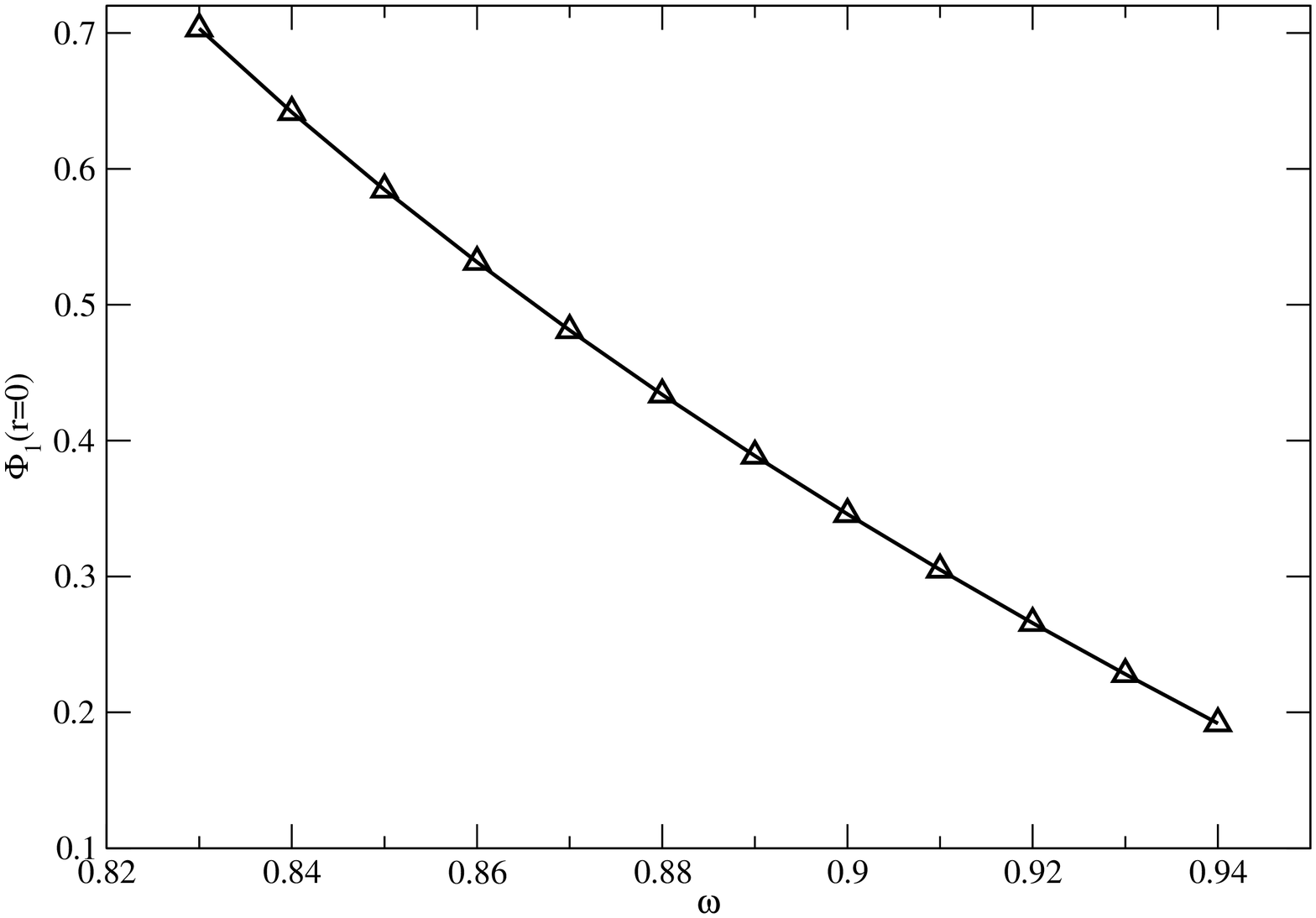}
\includegraphics[width=.5\textwidth]{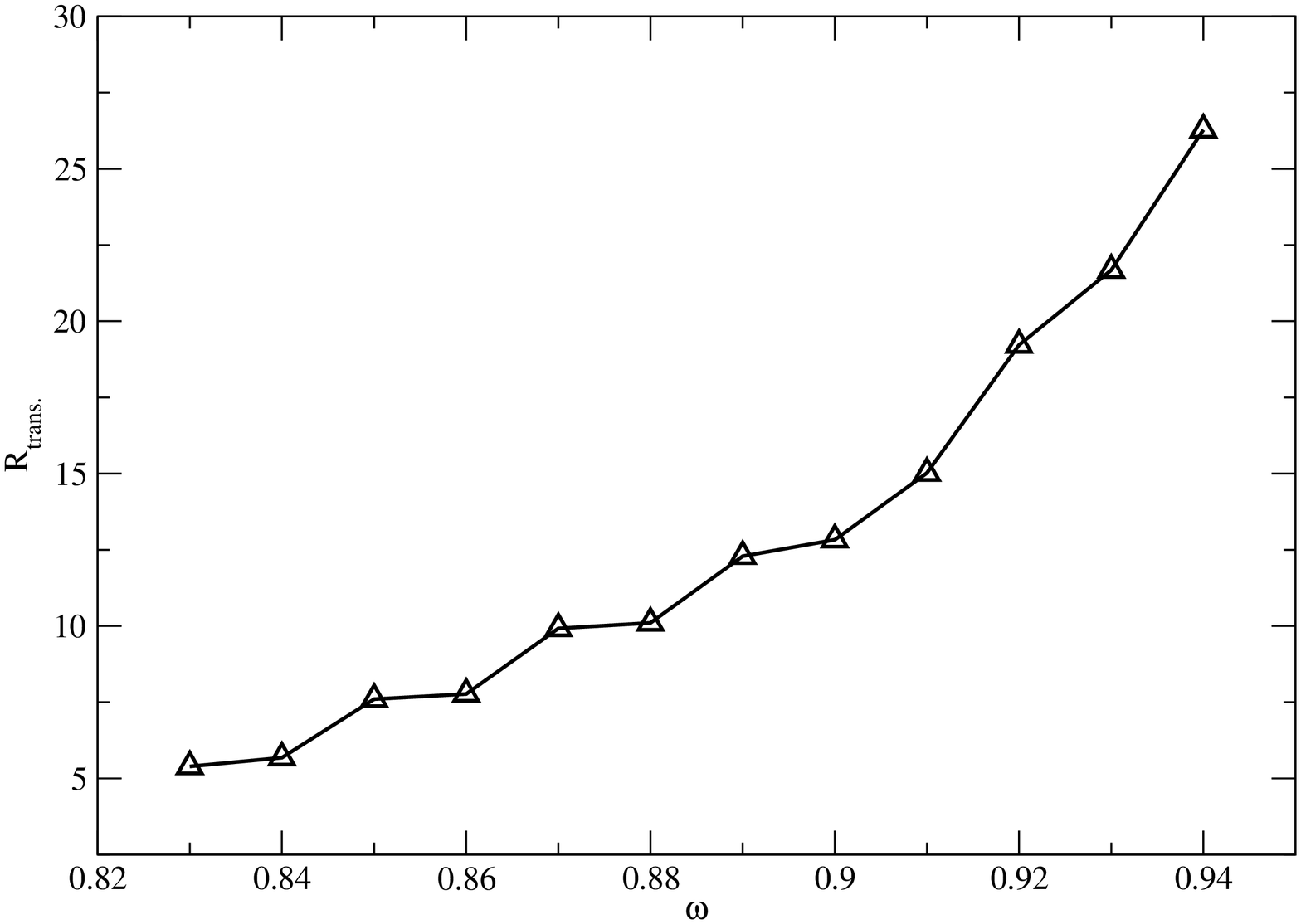}
\caption{ \label{f:global}
The first panel shows the total mass of the system, the second one gives the value of $\Phi_1$ at the origin and the third one the transition radius defined as the smallest radius for which $\Phi_3 = 0$. All quantities are computed as a function of $\omega$, in the high resolution case.}
\end{figure}

Some global quantities are shown in the various panels of Fig. \ref{f:global}. The first panel shows the total mass of the system. Given the behavior of the masses shown in the first panel of Fig. \ref{f:conv}, we define the mass as the mean of the value given for $A$ and $B$, that is $\l(r_A + r_B\r)/4$. A maximum mass of  $M_{\text{max}}=0.60535$ is attained for $\omega_{\rm min} =0.8608$. This is consistent with the values given in \cite{LopezMB02,AlcubBGMNU03} where the authors found $\omega_{\rm min}= 0.864$ and $M_{\text{max}}=0.607$. The presence of this maximum is important because oscillatons with $\omega<\omega_{\text{min}}$ are unstable. The second panel of Fig. \ref{f:global}  gives the value of $\Phi_1$ at the origin. Finally, in the third panel, we show the transition radius $R_{\rm trans}$ defined at the radius at which the oscillatory tail begins to dominate $\Phi_3$. More precisely, we define $R_{\rm trans}$ as the first radius for which $\Phi_3$ vanishes. As expected $R_{\rm trans}$ increases with $\omega$. The wiggling of the curve is just an effect of what the phase of the oscillatory tail is when it starts to dominate.

\begin{figure}[htbp]
\includegraphics[width=.5\textwidth]{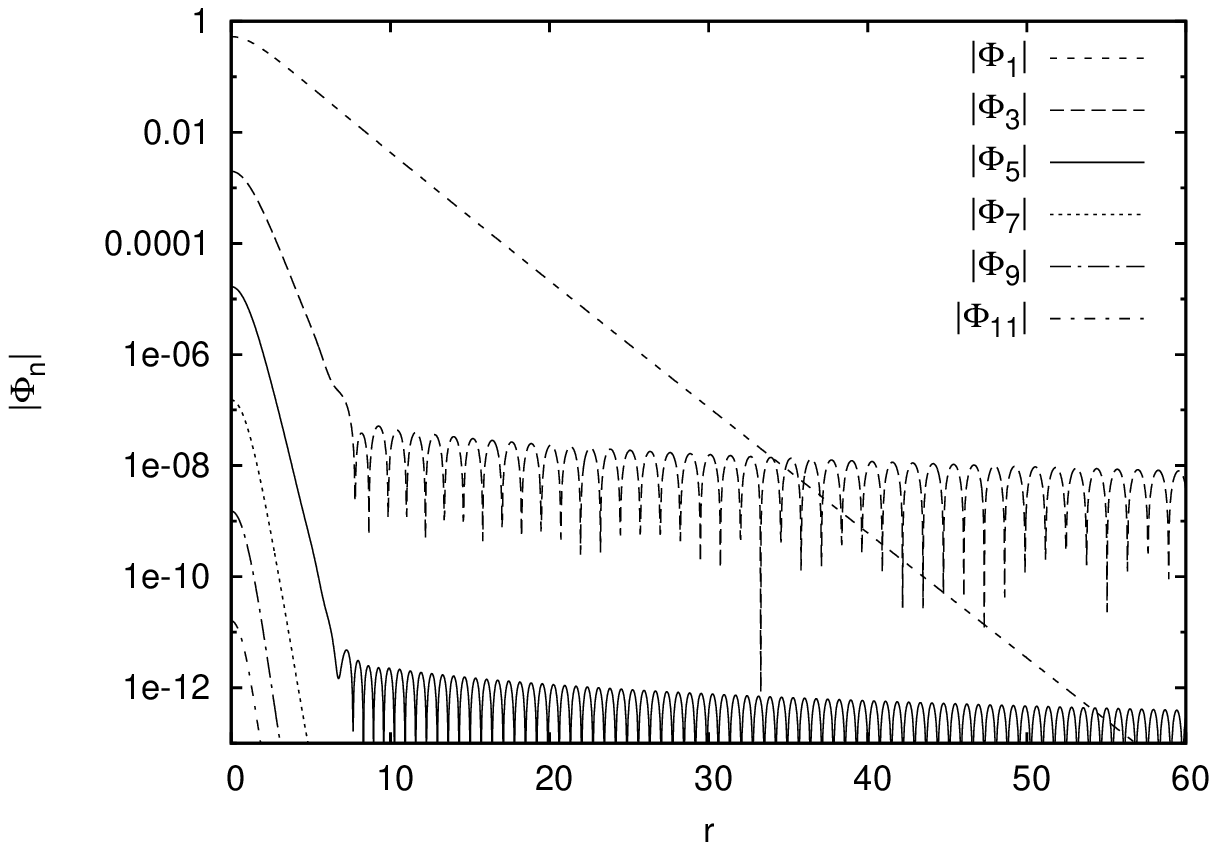}
\includegraphics[width=.5\textwidth]{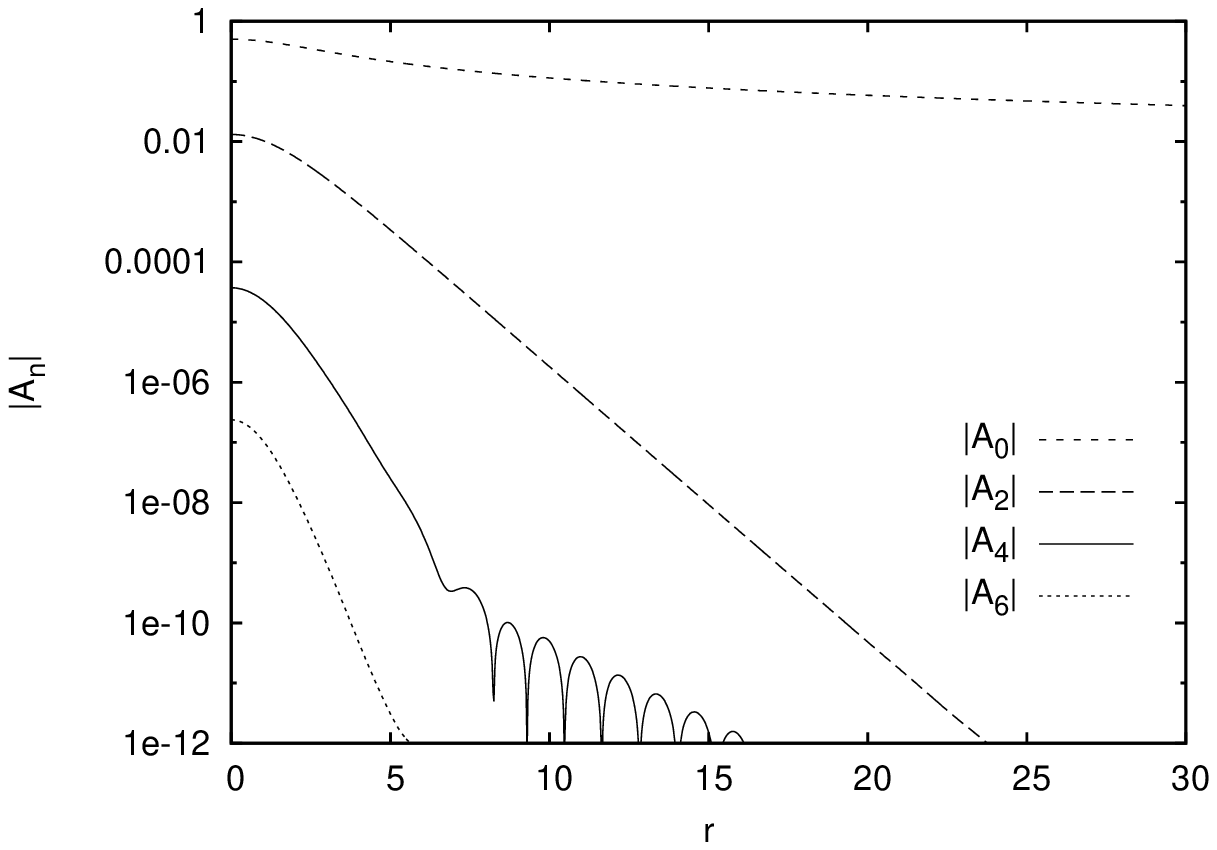}
\includegraphics[width=.5\textwidth]{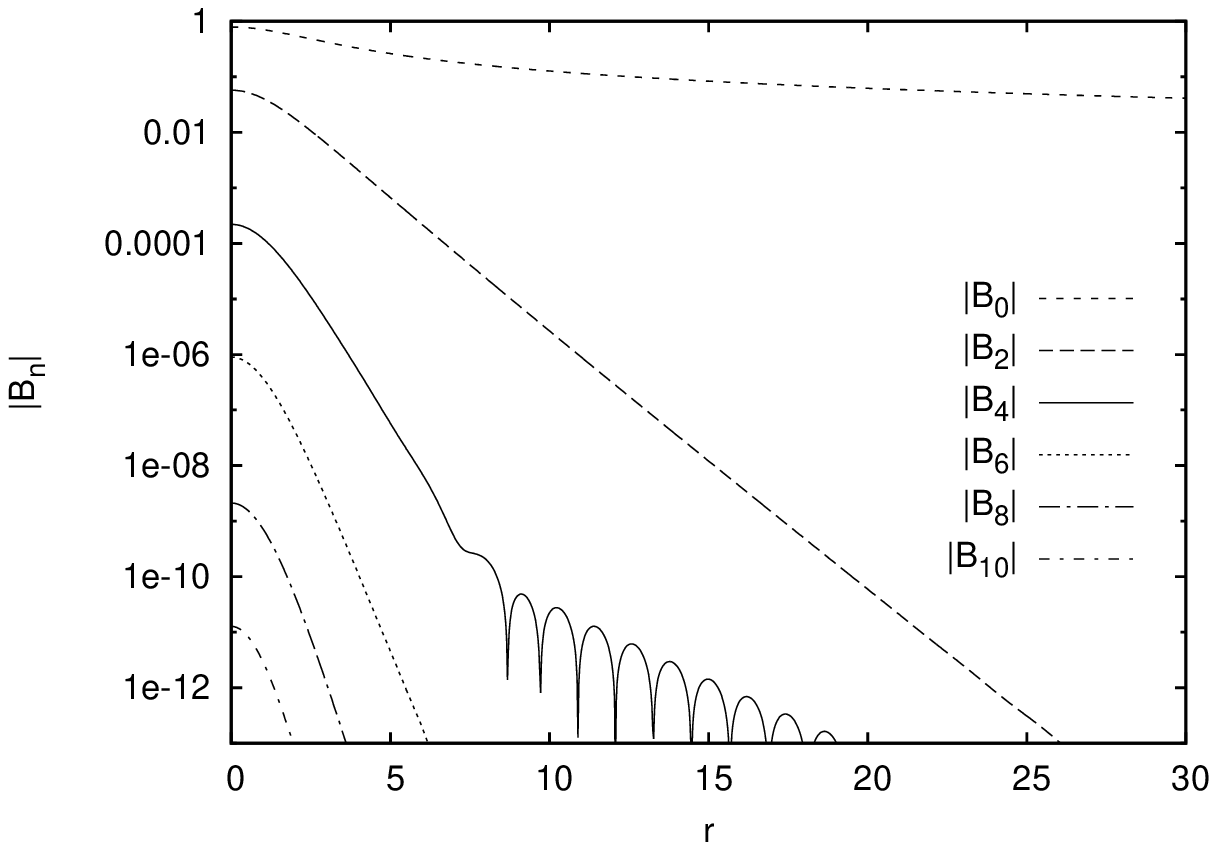}
\caption{ \label{f:fields}
Values of the various modes for $\Phi$ (first panel) and the metric fields $A$ and $B$ (second and third panels), for the oscillaton with maximum mass, corresponding to
$\omega= \omega_{\rm min} = 0.8608$.
}
\end{figure}

The various modes are shown as a function of the radius in Fig. \ref{f:fields} for the configuration of maximum mass (i.e. for $\omega=\omega_{\rm min}$). The dominating oscillatory term in $\Phi_3$ is clearly visible. The oscillations that can be seen on the metric fields come solely from the coupling with the scalar field. Indeed, remember that no matching with oscillatory tails is used for the metric fields, as explained in Sec. \ref{ss:matching}. When we tried comparing the values of $\Phi_n$ with the plots of \cite{Lopez02, AlcubBGMNU03} (for the appropriate $\omega$), we observed several orders of magnitude difference for $n>1$ whereas the dominating mode $n=1$ is in relatively good agreement. We have currently no explanation for this discrepancy but feel rather confident with our results, given the various tests exhibited in Sec. \ref{ss:tests}. As far as the metric fields are concerned, a direct comparison with \cite{Lopez02, AlcubBGMNU03} is not easy because of the use of different systems of coordinates. Nevertheless, we are puzzled by the fact that the metric fields shown in \cite{Lopez02, AlcubBGMNU03}, at least for $n>0$, do not seem to be even at the origin, as they should. It might be an effect of the resolution of the plots however.

The numerical values of various quantities are given in Table \ref{t:numerical}.
\begin{table}

\centering
\caption[]{\label{t:numerical}
Values of $\Phi_1$ at the origin, the amplitude of the dominating tail $C_3$ and the phase at which the minimum is attained $\delta_3$. The values are given as a function of $\omega$ and come from the computation with the highest resolution.
}
\begin{tabular}{| c | c |  c | c  | c | }
\hline
$\omega$ & Mass & $\Phi_1\l(r=0\r)$ & $C_3$ & $\delta_3$ \\
\hline
$0.83$ & $0.59446$ & $0.70326$ & $6.5894 \cdot 10^{-6}$ & $+0.58$ \\
$0.84$ & $0.60036$ & $0.64172$ & $2.9848 \cdot 10^{-6}$ & $+0.15$ \\
$0.85$ & $0.60399$ & $0.58468$ & $1.2793 \cdot 10^{-6}$ & $-0.28$ \\
$0.855$ & $0.60495$ & $0.55760$ & $8.1692 \cdot 10^{-7}$ & $-0.50$ \\
$0.8575$ & $0.60522$ & $0.54438$ & $6.4825 \cdot 10^{-7}$ & $-0.61$ \\
$0.85875$ & $0.60530$ & $0.53785$ & $5.7641 \cdot 10^{-7}$ & $-0.67$ \\
$0.86$ & $0.60535$ & $0.53137$ & $5.1183 \cdot 10^{-7}$ & $-0.72$ \\
$0.860625$ & $0.60535$ & $0.52815$ & $4.8210 \cdot 10^{-7}$ & $-0.75$ \\
$0.86125$ & $0.60535$ & $0.52494$ & $4.5392 \cdot 10^{-7}$ & $-0.78$ \\
$0.861875$ & $0.60534$ & $0.52174$ & $4.2724 \cdot 10^{-7}$ & $-0.81$ \\
$0.8625$ & $0.60532$ & $0.51855$ & $4.0201 \cdot 10^{-7}$ & $-0.83$ \\
$0.865$ & $0.60515$ & $0.50593$ & $3.1399 \cdot 10^{-7}$ & $-0.94$ \\
$0.87$ & $0.60437$ & $0.48122$ & $1.8815 \cdot 10^{-7}$ & $-1.17$ \\
$0.88$ & $0.60093$ & $0.43380$ & $6.2311 \cdot 10^{-8}$ & $+1.53$ \\
$0.89$ & $0.59484$ & $0.38878$ & $1.8127 \cdot 10^{-8}$ & $+1.07$ \\
$0.9$ & $0.58588$ & $0.34589$ & $4.4800 \cdot 10^{-9}$ & $+0.62$ \\
$0.91$ & $0.57371$ & $0.30493$ & $8.9842 \cdot 10^{-10}$ & $+0.16$ \\
$0.92$ & $0.55792$ & $0.26570$ & $1.3691 \cdot 10^{-10}$ & $-0.28$ \\
$0.93$ & $0.53796$ & $0.22805$ & $1.4368 \cdot 10^{-11}$ & $-0.72$ \\
$0.94$ & $0.51310$ & $0.19187$ & $8.8780 \cdot 10^{-13}$ & $-1.15$ \\
\hline
\end{tabular}
\end{table}

\section{Small-amplitude expansion}
\label{s:small}
\subsection{Review of the formalism}

Oscillatons in the small amplitude limit can be well described by an asymptotic expansion in terms of a small-amplitude parameter \cite{FodorFM10}. At leading order, configurations with any frequency can be characterized by the localized solution of a pair of ordinary differential equations. Since we wish to compare the numerical results in Sec. \ref{s:numerical} with those obtained by the small-amplitude expansion we give a short review of the main definitions and results in \cite{FodorFM10}. As the amplitude of oscillatons decreases, the geometry approaches the Minkowski metric, and the frequency approaches the mass limit from below, which has been set to $1$ in our case. At the same time, the size of oscillatons grows without limit, as is also indicated by the asymptotic behavior (\ref{e:shphi1}) of the leading mode of the scalar field. This motivates the introduction of the rescaled radial coordinate
\begin{equation}
\rho=\varepsilon r \ .
\end{equation}
The relation between the the small-amplitude parameter $\varepsilon$ and the fundamental frequency $\omega$ can be chosen as
\begin{equation}
\varepsilon^2=1-\omega^2 \ .
\end{equation}
The field $\Phi$ and the metric functions are then expanded in even powers of $\varepsilon$ as
\begin{align}
\Phi&=\sum_{k=1}^\infty\epsilon^{2k}\tilde\Phi_{2k} \,,\\
A&=1+\sum_{k=1}^\infty\epsilon^{2k}\tilde A_{2k} \,, \\
B&=1+\sum_{k=1}^\infty\epsilon^{2k}\tilde B_{2k} \,,
\end{align}
where we used tilde notation to distinguish the coefficients of the $\varepsilon$ expansion from the Fourier components defined in \eqref{e:modeA}-\eqref{e:modeP}. To leading order the field $\Phi$ turns out to be proportional to $\varepsilon^2$. From the assumption that the functions remain bounded as time passes follows that the configuration has to be periodic, and the time dependence of the coefficients can be integrated out. Writing out the first couple of orders,
\begin{align}
\Phi&=\varepsilon^2p_2\cos(\omega t)
+\varepsilon^4p_4\cos(\omega t)+\varepsilon^6p_6\cos(\omega t)+\varepsilon^6\left(
\frac{3p_2^3}{128}+\frac{p_2a_4^{(2)}}{8}\right)\cos(3\omega t)
+\mathcal{O}(\varepsilon^8) \,,  \label{e:phieps}\\
A&=1+\varepsilon^2a_2+\varepsilon^4\left[
a_4^{(0)}+a_4^{(2)}\cos(2\omega t)\right]+\mathcal{O}(\varepsilon^6)
\,, \\
B&=1-\varepsilon^2 a_2+\varepsilon^4\left[
b_4-\frac{p_2^2}{8}\cos(2\omega t)
\right]+\mathcal{O}(\varepsilon^6)
\,,
\end{align}
where $p_2$, $p_4$, $p_6$, $a_2$, $a_4^{(0)}$, $a_4^{(2)}$ and $b_4$ are functions of the radial coordinate $\rho$. These can be obtained order by order by solving ordinary differential equations arising in the small-amplitude expansion. The leading order configuration is determined by the time-independent Schr\"odinger-Newton equations
\begin{align}
\frac{d^2S}{d\rho^2}
+\frac{2}{\rho}\,\frac{d S}{d\rho}
+s S&=0
\,, \label{Seq}\\
\frac{d^2s}{d\rho^2}
+\frac{2}{\rho}\,\frac{d s}{d\rho}
+S^2&=0
\,, \label{seq}
\end{align}
where the functions $s$ and $S$ are defined by
\begin{equation}
s=-1-a_2 \ , \quad S=p_2\sqrt{2} \ .
\label{eq:sands}
\end{equation}

The total mass, $M=2r_0$, of the configuration can be expanded as
\begin{equation}
M=\varepsilon M_1
+\varepsilon^{3}M_2
+\mathcal{O}\left(\varepsilon^{5}\right) , \label{eq:totmass}
\end{equation}
where the numerical values of the constants are $M_1=1.75266$ and $M_2=-2.11742$.

The small-amplitude expansion gives exponentially localized configurations to all orders, and consequently it cannot describe the oscillating tail responsible for the mass loss of oscillatons. Although the small-amplitude expansion is an asymptotic one, it gives a useful representation of the core region. The dominant radiating mode \eqref{e:shphin} in $\Phi_3$ can be calculated to leading order by the extension of the Fourier mode equations into the complex $r$ plane, and employing Borel summation in a region around the resulting pole \cite{FodorFM10}, giving
\begin{equation}
C_3=\frac{k\pi Q }{\varepsilon}\exp\left(-\frac{\sqrt{8}Q}{\varepsilon}\right) , \label{eq:c3epskq}
\end{equation}
where $Q$ is the distance of the pole from the real axis of the solution of the Schr\"odinger-Newton equations, and $k$ is a constant. Numerically $Q=3.97736$, $k=0.301$ and
\begin{equation}
C_3=\frac{3.761}{\varepsilon}\exp\left(-\frac{11.2497}{\varepsilon}\right) \ .  \label{eq:c3eps}
\end{equation}

\subsection{Comparison with the numerical solution of the Fourier mode equation}

\begin{figure}[htbp]
\includegraphics[width=.5\textwidth]{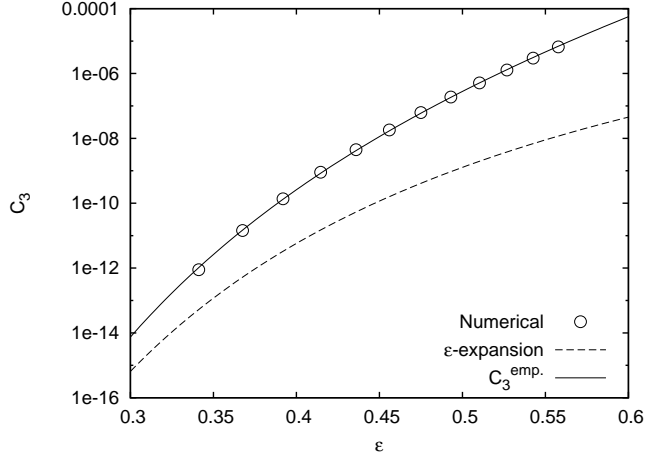}
\caption{ \label{f:c3cmp}
Comparison of the numerical and theoretical values of $C_3$. A very good empirical fit, given by \eqref{eq:c3emp} is also shown.}
\end{figure}

In Fig.~\ref{f:c3cmp} we again plot the highest resolution tail amplitude data of $\Phi_3$ from Fig.~\ref{f:c3}, this time as a function of the small-amplitude parameter $\varepsilon=\sqrt{1-\omega^2}$. The theoretical curve from \eqref{eq:c3eps} goes well below the data points.
The following empirical formula
\begin{equation}
C_3^{\text{emp.}}=\frac{3.761}{\varepsilon}\left(1+\varepsilon^2\right)^{16.63}\exp\left[-\frac{11.2497}{\varepsilon}\left(1-0.2990\,\varepsilon^2\right)\right] \ ,  \label{eq:c3emp}
\end{equation}
gives an excellent fit to the $C_3$ data points, moreover, for small $\varepsilon$ it approaches the theoretical result \eqref{eq:c3eps}. A natural interpretation of this result is that the number $Q$ in the exponent of \eqref{eq:c3epskq}, describing the distance of the pole from the real axis, has a polynomial $\varepsilon$ dependence, and the constant $k$ also changes for large $\varepsilon$.

\begin{figure}[htbp]
\includegraphics[width=.5\textwidth]{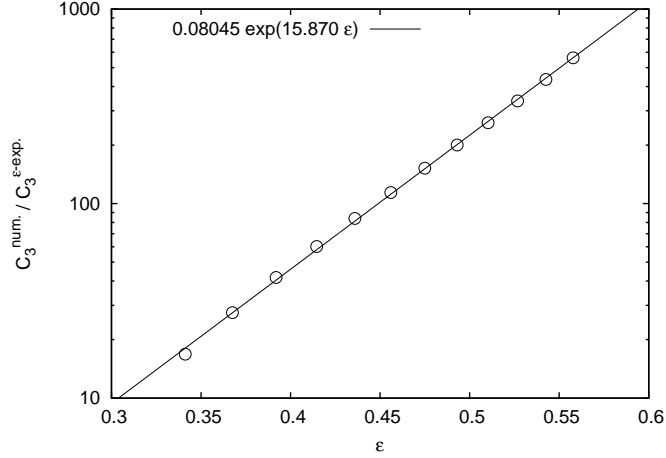}
\caption{ \label{f:c3rat}
Ratio of the numerical and theoretical values of $C_3$, together with a fit of an exponential curve through the data points.}
\end{figure}

The ratio of the numerically and theoretically obtained tail amplitudes is plotted in Fig.~\ref{f:c3rat}.
The numerical value is several hundred times bigger for large $\varepsilon$ but the ratio decreases exponentially when $\varepsilon$ gets smaller. Even for the smallest $\varepsilon$ where the tail could be calculated, with amplitude of about $10^{-12}$, the difference is about twentyfold. Although because of numerical errors we cannot calculate the tail for smaller $\varepsilon$, it appears convincing that the two method would converge for even smaller amplitudes. It can be seen in Fig.~\ref{f:c3rat} that the fit of the exponential curve
\begin{equation}
\frac{C_3^{\text{num.}}}{C_3^{\varepsilon\text{-exp.}}}=0.08045\exp(15.870\,\varepsilon)
\end{equation}
to the ratio of the numerical and analytical result gives a very good approximation for $0.3<\varepsilon<0.6$, but is clearly not appropriate for smaller $\varepsilon$ values, since in the $\varepsilon\to 0$ limit it does not approach the value $1$.

\begin{figure}[htbp]
\includegraphics[width=.5\textwidth]{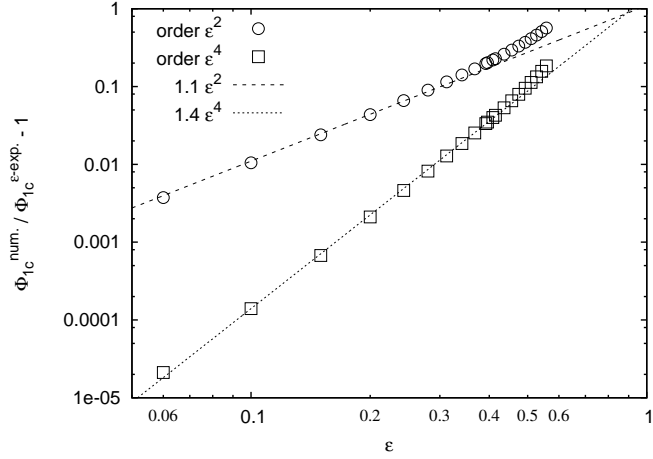}
\caption{ \label{f:p1cmp}
Relative difference of the numerically calculated central value of the $\Phi_1$ Fourier mode and the one calculated by the $\varepsilon$ expansion method, using the $\varepsilon^2$ and $\varepsilon^4$ order approximations from \eqref{e:phi1c}. To show the tendency of the points two curves proportional to $\varepsilon^2$ and $\varepsilon^4$ are also plotted.}
\end{figure}

The reason for the large difference between the theoretical and numerical tail amplitudes can be better understood by inspecting how precisely the core region is described by the small-amplitude expansion when $\varepsilon$ goes above $0.34$. To do this, we compare the Fourier components of $\Phi$ at the origin. From \eqref{e:phieps} it follows that $\Phi_1=\varepsilon^2 p_2+\varepsilon^4 p_4+\mathcal{O}(\varepsilon^6)$. Substituting the numerical values of $p_2$ and $p_4$, for the central value of the leading Fourier component we obtain
\begin{equation}
\Phi_{1c}^{\varepsilon\text{-exp.}}=1.44461\,\varepsilon^2+1.49305\,\varepsilon^4+\mathcal{O}(\varepsilon^6) \ . \label{e:phi1c}
\end{equation}
In Fig.~\ref{f:p1cmp} we plot the relative difference of the numerical and the analytical results for $\Phi_{1c}$. Here we can include numerical values even for $\varepsilon<0.34$, since the core region can be reliably calculated by numerically solving the Fourier mode equations even if the tail amplitude in $\Phi_3$ goes below the numerical noise.
We can see that the first two terms given by the small-amplitude expansion \eqref{e:phi1c} give excellent approximation to the central amplitude for low and moderate $\varepsilon$ values. For the largest $\varepsilon$ the difference between the numerical and theoretical values grows to about $20$ percent. At $\varepsilon=0.34$ it decreases to about $2$ percent.

\begin{figure}[htbp]
\includegraphics[width=.5\textwidth]{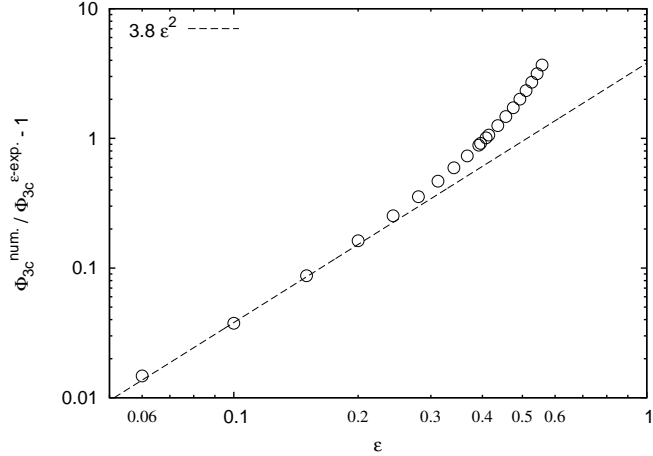}
\caption{ \label{f:p3cmp}
Relative difference of the numerically calculated central value of the $\Phi_3$ Fourier mode and the one calculated by the $\varepsilon$ expansion method. To show the tendency a curve proportional to $\varepsilon^2$ is also plotted.}
\end{figure}

The error of the $\varepsilon$ expansion for $\varepsilon>0.34$ is considerably larger for the next Fourier component $\Phi_3$, which component also gives the dominant contribution to the oscillating tail responsible for the energy loss. Using \eqref{e:phieps} the small-amplitude expansion gives the central value
\begin{equation}
\Phi_{3c}^{\varepsilon\text{-exp.}}=0.0343467\,\varepsilon^6+\mathcal{O}(\varepsilon^8) \ . \label{e:phi3c}
\end{equation}
In Fig.~\ref{f:p3cmp} the relative difference of the numerical value of $\Phi_{3c}$ with respect to that given by \eqref{e:phi3c} is shown.
Here, for the largest parameter value, at $\varepsilon=0.5578$, the numerically calculated $\Phi_{3c}$ is about five times the one given by the small-amplitude expansion. For $\varepsilon=0.34$ the numerical value is still about $50\%$ bigger. Since the result \eqref{eq:c3epskq} is based on the assumption that the core is described by the small-amplitude expansion, this big difference in the central value of $\Phi_3$ makes the large error in the tail amplitude for $\varepsilon>0.34$ comprehensible.

\begin{figure}[htbp]
\includegraphics[width=.5\textwidth]{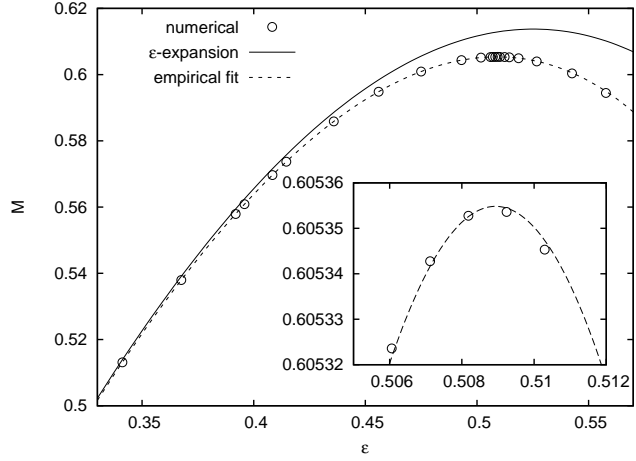}
\caption{ \label{f:masseps}
Numerically and analytically obtained masses of configurations with various $\varepsilon$ parameters. A quite precise empirical fit given by \eqref{eq:massfit} is also shown.}
\end{figure}

In Fig.~\ref{f:masseps} we plot the total mass of the oscillaton as a function of the $\varepsilon$ parameter. The first two terms in the small-amplitude expansion, given by \eqref{eq:totmass}, give a reasonable approximation, but the position and the value of the maximum is not extremely precise. This not so surprising because the maximum is reached for $\varepsilon \approx 0.5$ which is not particularly small.
It is possible to make a very good empirical fit of the form
\begin{equation}
M^{\text{emp.}}=\varepsilon M_1
+\varepsilon^{3}M_2
+\varepsilon^{5}M_3
+\varepsilon^{7}M_4
+\varepsilon^{9}M_5 \ , \label{eq:massfit}
\end{equation}
where $M_1$ and $M_2$ was kept at the value $M_1=1.75266$ and $M_2=-2.11742$ given by the small-amplitude expansion, and the fitted constants are $M_3=-0.24723$, $M_4=1.1749$ and $M_5=-4.1308$. The numerically obtained value for the place of the maximum is $\varepsilon_{\text{max}}=0.509$, corresponding to the frequency $\omega_{\text{min}}=0.8608$, and the maximal mass is $M_{\text{max}}=0.60535$. The importance of these considerations lies in the fact that oscillatons with $\varepsilon>\varepsilon_{\text{max}}$, or equivalently, $\omega<\omega_{\text{min}}$ are unstable. The value given in \cite{LopezMB02,AlcubBGMNU03} for $\omega_{\text{min}}$ is $0.864$ and for $M_{\text{max}}$ is $0.607$.
Reintroducing the scalar field mass, $m$, and expressing it in units currently used in Particle Physics, i.e.\ in eV$/mc^2$, we obtain the maximal oscillaton mass in kg-s:
\begin{equation}
 M_{\text{max}}=1.6085\times 10^{20}\text{kg}\frac{\text{eV}}{mc^2}\,.
\end{equation}

\section{Mass loss rate}
\label{s:longevity}

In order to calculate the rate by which the mass M of the oscillaton decreases, we have to calculate the mass-energy carried out by the spherical scalar wave
\begin{equation}
\Phi=\frac{C_3}{r}\cos\left(\lambda_3 r-3\omega t\right) , \label{e:outwave}
\end{equation}
where $\lambda_3=\sqrt{9\omega^2-1}$. Restoring the $1/\sqrt{8\pi}$ factors from \eqref{e:scscale} into the scalar field, the mass-energy carried by the wave is
\begin{equation}
\frac{dM}{dt}=\frac{r^2}{2}\,\frac{d\Phi}{dt}\,\frac{d\Phi}{dr} \ .
\end{equation}
Substituting \eqref{e:outwave}, taking the large $r$ limit, and averaging in time
\begin{equation}
\frac{dM}{dt}=-\frac{3}{4}C_3^2\omega\sqrt{9\omega^2-1} \ . \label{e:dmdt}
\end{equation}
Using the empirical formula \eqref{eq:c3emp} for $C_3$, for the mass-loss rate we obtain
\begin{equation}
\frac{dM}{dt}=-10.61\frac{\omega\sqrt{9\omega^2-1}}{\varepsilon^2}\left(1+\varepsilon^2\right)^{33.26}\exp\left[-\frac{22.4993}{\varepsilon}\left(1-0.2990\,\varepsilon^2\right)\right] \ , \label{e:dmdtemp}
\end{equation}
where $\omega=\sqrt{1-\varepsilon^2}$. Using the version of \eqref{e:dmdt} with $\omega=1$, and substituting $C_3$ from \eqref{eq:c3eps}, we get the leading order small-amplitude result already obtained in \cite{FodorFM10}, 
\begin{equation}
\frac{dM}{dt}=-\frac{30.0}{\varepsilon^2}\exp\left(-\frac{22.4993}{\varepsilon}\right) \ . \label{e:dmdteps}
\end{equation}
Although for small $\varepsilon$ the empirical mass-loss rate formula \eqref{e:dmdtemp} can be approximated by \eqref{e:dmdteps}, for close to maximal amplitude oscillatons \eqref{e:dmdtemp} gives significantly higher radiation loss.

In order to be able to draw physical conclusions, we have to restore the scalar field mass $m$ into the expressions. This can be done by substituting $t\to mt$, $r\to mr$ into the equations (see e.g.~\cite{FodorFM10}). Then we have to substitute $M\to mM$, so the right hand sides of \eqref{eq:totmass} and \eqref{eq:massfit} will get a $1/m$ factor. The $m$ factors drop out from \eqref{e:dmdt}, \eqref{e:dmdtemp} and \eqref{e:dmdteps}, so they remain valid for any scalar field mass $m$. The physical frequency of the oscillaton will be $\tilde\omega=m\omega=m\sqrt{1-\varepsilon^2}$.

Taking \eqref{e:dmdtemp} at $\varepsilon_{\text{max}}=0.509$ corresponding to the maximal mass $M_{\text{max}}=0.60535$ and dividing by $M_{\text{max}}$ we get the relative mass loss rate for the heaviest stable oscillaton:
\begin{equation}
\left(\frac{1}{M}\frac{dM}{dt}\right)_{M=M_{\text{max}}}=-5.917\times 10^{-13} m \ ,
\end{equation}
where $m$ is the scalar field mass in Planck units. This is about $14000$ times larger than the theoretical estimation $-4.3\times 10^{-17}$ in \cite{FodorFM10} obtained from the leading order small-amplitude behavior. The large difference arises because the small-amplitude analysis underestimates the amplitude of the radiating tail by about a factor of $100$ for $\varepsilon$ as large as $\varepsilon_{\text{max}}=0.509$.

It is natural to start with a maximal mass configuration with $M=M_{\mathrm{max}}$, and ask for the time period until the mass decreases to a certain part of its original value. Since the elapsed time $t$ is inversely proportional to the scalar field mass $m$, in table \ref{ttable} we list the product $tm$.
\begin{table}[htbp]
\begin{tabular}{|l|l|l|l|l|l|l|}
\hline
& \multicolumn{3}{|c|}{$\varepsilon$-expansion}
& \multicolumn{3}{|c|}{Fourier expansion}\\
\cline{2-7}
$\frac{M_{\mathrm{max}}-M}{M_{\mathrm{max}}}$ & $\varepsilon$
  & $tm$
  & $\frac{t}{\mathrm{year}}\,\frac{mc^2}{eV}$  & $\varepsilon$
  & $tm$
  & $\frac{t}{\mathrm{year}}\,\frac{mc^2}{eV}$\\
\hline
$0.01$ & $0.482$ & $3.45\cdot 10^{15}$ & $7.21\cdot 10^{-8}$ & $0.469$ & $7.03\cdot 10^{11}$ & $1.46\cdot 10^{-11}$ \\
$0.1$ & $0.383$ & $8.65\cdot 10^{20}$ & $0.0180$ & $0.376$ & $2.63\cdot 10^{18}$ & $5.48\cdot 10^{-5}$ \\
$0.2$ & $0.320$ & $6.69\cdot 10^{25}$ & $1400$ & $0.314$ & $1.16\cdot 10^{24}$ & $24.2$ \\
$0.3$ & $0.267$ & $2.55\cdot 10^{31}$ & $5.32\cdot 10^{8}$ & $0.264$ & $1.71\cdot 10^{30}$ & $3.56\cdot 10^{7}$ \\
$0.4$ & $0.224$ & $2.74\cdot 10^{38}$ & $5.72\cdot 10^{15}$ & $0.220$ & $5.87\cdot 10^{37}$ & $1.23\cdot 10^{15}$ \\
$0.5$ & $0.182$ & $9.54\cdot 10^{47}$ & $1.99\cdot 10^{25}$ & $0.180$ & $5.98\cdot 10^{47}$ & $1.25\cdot 10^{25}$ \\
$0.6$ & $0.144$ & $1.10\cdot 10^{62}$ & $2.29\cdot 10^{39}$ & $0.142$ & $2.03\cdot 10^{62}$ & $4.24\cdot 10^{39}$ \\
$0.7$ & $0.107$ & $1.78\cdot 10^{85}$ & $3.72\cdot 10^{62}$ & $0.105$ & $1.18\cdot 10^{86}$ & $2.45\cdot 10^{63}$ \\
\hline
\end{tabular}
\caption{\label{ttable}
The time necessary for an initially maximal mass oscillaton to lose the given part of its mass. The value of $tm$ is given in Planck units, and also when the time is measured in years and the scalar mass in $eV/c^2$ units. The values calculated by the small-amplitude expansion method and by the numerical solution of the Fourier mode equations are also given.
}
\end{table}
We give the elapsed time calculated first by using the small-amplitude expansion results and then by the more precise Fourier mode decomposition method. In the first case we calculate $\frac{dM}{dt}$ from \eqref{e:dmdteps}, and approximate the $\varepsilon$ dependence of the mass by \eqref{eq:totmass}. In order to obtain the more precise result we calculate the mass-loss rate from the empirical formula \eqref{e:dmdtemp}, and approximate the mass function by \eqref{eq:massfit}. Unfortunately, a sign mistake was made in \cite{FodorFM10} when substituting the numerical value of $M^{(2)}$ during the calculation of the numbers in Table VII and VIII, which resulted in smaller radiation rate for large amplitude oscillatons than the correct rate. This is also the reason why we also present the numbers obtained by the $\varepsilon$ expansion here.

Next we address the question that how much of its mass an initially maximal mass oscillaton loses during a time period corresponding to the age of the universe, which we take to be $1.37\cdot 10^{10}$ years. In Table \ref{mstable} we list the resulting oscillaton masses in units of solar masses $(M_\odot)$ as a function of the scalar field mass in $eV/c^2$ units.
\begin{table}[htbp]
\begin{tabular}{|l|l|l|l|l|l|l|}
\hline
& \multicolumn{3}{|c|}{$\varepsilon$-expansion}
& \multicolumn{3}{|c|}{Fourier expansion}\\
\cline{2-7}
$\frac{mc^2}{eV}$ & $\varepsilon_{\mathrm{max}}-\varepsilon$
  & $\frac{M}{M_\odot}$
  & $\frac{M_{\mathrm{max}}-M}{M_{\mathrm{max}}}$ & $\varepsilon_{\mathrm{max}}-\varepsilon$
  & $\frac{M}{M_\odot}$
  & $\frac{M_{\mathrm{max}}-M}{M_{\mathrm{max}}}$ \\
\hline
$10^{-35}$ & $2.31\cdot 10^{-10}$
   & $8.20\cdot 10^{24}$ & $2.91\cdot 10^{-19}$ & $2.37\cdot 10^{-8}$
   & $8.09\cdot 10^{24}$ & $3.75\cdot 10^{-15}$ \\
$10^{-30}$ & $7.31\cdot 10^{-8}$
   & $8.20\cdot 10^{19}$ & $2.91\cdot 10^{-14}$ & $7.50\cdot 10^{-6}$
   & $8.09\cdot 10^{19}$ & $3.74\cdot 10^{-10}$ \\
$10^{-25}$ & $2.31\cdot 10^{-5}$
   & $8.20\cdot 10^{14}$ & $2.90\cdot 10^{-9}$ & $2.18\cdot 10^{-3}$
   & $8.09\cdot 10^{14}$ & $3.16\cdot 10^{-5}$ \\
$10^{-20}$ & $0.00621$
   & $8.20\cdot 10^{9}$ & $2.09\cdot 10^{-4}$ & $0.0544$
   & $7.94\cdot 10^{9}$ & $0.0183$ \\
$10^{-15}$ & $0.0883$
   & $7.87\cdot 10^{4}$ & $0.0400$ & $0.126$
   & $7.36\cdot 10^{4}$ & $0.0896$ \\
$10^{-10}$ & $0.169$
   & $7.06\cdot 10^{-1}$ & $0.139$ & $0.183$
   & $6.65\cdot 10^{-1}$ & $0.178$ \\
$10^{-5}$ & $0.226$
   & $6.24\cdot 10^{-6}$ & $0.238$ & $0.227$
   & $5.96\cdot 10^{-6}$ & $0.263$ \\
$1$ & $0.267$
   & $5.56\cdot 10^{-11}$ & $0.322$ & $0.262$
   & $5.36\cdot 10^{-11}$ & $0.337$ \\
$10^{5}$ & $0.298$
   & $4.98\cdot 10^{-16}$ & $0.392$ & $0.289$
   & $4.85\cdot 10^{-16}$ & $0.401$ \\
$10^{10}$ & $0.323$
   & $4.51\cdot 10^{-21}$ & $0.450$ & $0.311$
   & $4.41\cdot 10^{-21}$ & $0.454$ \\
$10^{15}$ & $0.342$
   & $4.11\cdot 10^{-26}$ & $0.499$ & $0.329$
   & $4.04\cdot 10^{-26}$ & $0.500$ \\
\hline
\end{tabular}
\caption{\label{mstable}
  Mass $M$ of an initially maximal mass oscillaton after a period
  corresponding to the age of the universe for various scalar
  field masses. The decrease in $\varepsilon$ from
  $\varepsilon_{\mathrm{max}}$, and the relative mass change
  $(M_{\mathrm{max}}-M)/M_{\mathrm{max}}$ is also given. At the small-amplitude expansion $\varepsilon_{\mathrm{max}}=0.525$ is used, while the mode decomposition value is $\varepsilon_{\mathrm{max}}=0.509$.
}
\end{table}

\section{Conclusions}

We have presented a rather detailed numerical study of the structure of spherically symmetric, {\sl time periodic} oscillaton solutions of the Einstein-Klein-Gordon equations. 
We solved the equations by using a two-dimensional spectral method for both the radial coordinate and time. The use of spectral methods enabled us to obtain very precise solutions at a moderate computational cost.

For the first time we have succeeded in computing the amplitude of the standing wave tail of the time periodic oscillatons. The amplitude of those tails has been found to be very small indeed as compared to the central amplitude of an oscillaton. This implies that truly localized, time-periodic, asymptotically flat oscillatons do not exist, rather oscillatons of finite mass created by physical processes continuously lose some of their mass due to scalar radiation. It should be noted, however, that since the radiation rate of oscillatons decreases sufficiently rapidly as their total mass decreases, oscillatons cannot radiate away their mass in a finite time.

Using the precise numerical results we have derived a semi-empirical mass loss formula 
of ``physical'' oscillatons of finite mass. The results show that the previous computations of the mass loss rates in the small amplitude limit underestimated the true rate by several orders of magnitude for larger amplitude oscillatons.
Nevertheless the qualitative picture of an oscillon as a lump losing its mass extremely slowly prevails making these objects of physical interest, such as dark matter candidates.
The agreement with analytical results obtained in the limit of small amplitudes is very satisfactory as far as the structure of the core is concerned. Concerning the amplitude of the oscillatory tail it is more difficult to do a precise comparison. Indeed, the tail is very small and can be accurately extracted from the numerical simulations only when the amplitude of the oscillatons is not small, so in the region where the analytical approximation is expected to fail. The numerical and analytical results 
are coherent with each other.

We have also computed the value of the maximal mass which an oscillaton may have, 
together with the corresponding value of frequency.

\begin{acknowledgments}

This research has been supported by OTKA Grant No. NI68228.
\end{acknowledgments}

\end{document}